\documentclass[twocolumn,trackchanges]{aastex7}
\usepackage{tabularx}
\usepackage{array}
\usepackage{ragged2e}
\usepackage{comment}
\newcolumntype{Y}{>{\RaggedRight\arraybackslash}X}

\begin{document}

\title{Elemental Composition Evolution during the 2024 September 30 Solar Eruption: A Comparison of Hot and Cool Plasma Components with Solar Orbiter/SPICE, Hinode/EIS, and Chandrayaan-2/XSM}

\author[orcid=0000-0003-0583-0516,sname='Molnar']{Momchil E. Molnar}
\affiliation{Southwest Research Institute, Boulder, CO, 80302, USA}
\email[show]{momchil.molnar@swri.com}  

\author[orcid=0000-0001-7016-7226]{Joseph E. Plowman}
\affiliation{Southwest Research Institute, Boulder, CO, 80302, USA}
\email[noshow]{}

\author[orcid=0000-0001-8702-8273]{Amir Caspi}
\affiliation{Southwest Research Institute, Boulder, CO, 80302, USA}
\email[noshow]{}

\author[orcid=0000-0001-8504-2725]{Ritesh Patel}
\affiliation{Southwest Research Institute, Boulder, CO, 80302, USA}
\email[noshow]{}

\author[orcid=0000-0001-9035-3245]{Arpit Kumar Shrivastav}
\affiliation{Southwest Research Institute, Boulder, CO, 80302, USA}
\email[noshow]{}

\author[orcid=0000-0003-0256-9295]{Tania Varesano}
\affiliation{Ann and H.J. Smead Aerospace Engineering Sciences, University of Colorado Boulder, Boulder, CO, 80303, USA}
\affiliation{Southwest Research Institute, Boulder, CO, 80302, USA}
\email[noshow]{}

\author[orcid=0000-0001-8830-1200]{D. M. Hassler}
\affiliation{Southwest Research Institute, Boulder, CO, 80302, USA}
\email[noshow]{}

\author[orcid=0000-0001-9726-0738]{Ryan J. French}
\affiliation{Laboratory for Atmospheric and Space Physics, University of Colorado Boulder, Boulder, CO 80303, USA}
\email[noshow]{}

\author[orcid=0000-0002-7020-2826]{Biswajit Mondal}
\affiliation{University of Alabama in Huntsville, Huntsville, AL 35805, USA}
\affiliation{NASA Marshall Space Flight Center, ST12, Huntsville, AL 35812, USA}
\email[noshow]{}

\author[orcid=0000-0002-9270-6785]{L. P. Chitta}
\affiliation{Max-Planck-Institut f\"{u}r Sonnensystemforschung, 37077 G\"{o}ttingen, Germany}
\email[noshow]{}

\begin{abstract}
Solar plasma composition differs between the photosphere and corona over a range of timescales, with preferential enhancement of elements with low first ionization potential (FIP). 
However, the physical origin of the FIP fractionation remains incompletely understood. Furthermore, during flares, the FIP bias also exhibits rapid changes, associated with fast transport of material with different FIP biases. We present novel observations from Solar Orbiter SPICE and EUI, Hinode/EIS, and Chandrayaan-2 XSM instruments, finding rapid abundance changes in the emitting plasma, on timescales of minutes, during the eruptive M7.6-class solar flare observed on 2024 Sept 30. These instruments have wide temperature coverage and find contrasting abundance-evolution patterns  between the hotter and cooler plasma components. 3D reconstruction of the active region and additional observations from the Solar Orbiter STIX X-ray telescope show how the hot and cool plasma components, emitting in different spectral regions and observed with the various instruments, sample the plasma composition evolution in distinct locations within the observed flaring plasma. The bright post-flare loop tops observed by SPICE show coronal FIP bias, while the hot plasma observed with XSM exhibits FIP-bias decreasing from coronal to photospheric  during the impulsive phase. We interpret these observations as evidence of the X-ray diagnostics seeing hot coronal reconnection outflows mixing with chromospheric plasma as flare loops sequentially energize and relax, explaining why the FIP bias decreases from coronal to a hybrid; and the cool loop tops seen with SPICE show coronal abundances due to coronal material deposited near the looptops.
\end{abstract}

\section{Introduction}
\label{sec:Introduction}

The composition of the solar coronal plasma carries an imprint of 
its evolution that can constrain its origin -- the coronal plasma (and solar energetic particles) exhibit enhanced abundance of elements with low First Ionization Potential (FIP) \citep{1963ApJ...137..945P,1985ApJS...57..173M,1991AdSpR..11a.269M,1992ApJS...81..387F,2011ApJ...727L..13B}. The coronal enhancement of the abundances of elements 
with low First Ionization Potential (FIP) in the solar corona is observed to be 
a factor of $\sim$4 for most low-FIP elements (e.g., Fe, Ca, Si, Mg)~\citep{1992PhyS...46..202F}.
It is thought that this low-FIP enhancement in the corona is due to a fractionation process 
in the chromosphere~\citep{2004ApJ...614.1063L}. The degree of FIP fractionation is related to the 
time that solar plasma spends in the chromosphere, leading to open fields (where plasma can 
quickly propagate away) being associated with photospheric abundances and
closed features (where plasma is largely trapped) exhibiting coronal abundances \citep{Ko_2016, 2023ApJ...959...72M}. There are different ideas proposed to explain the fractionation, which are still under debate. These include theoretical models that rely on wave reflection from the transition region leading to a mean electric field, known as the effective ponderomotive force~\citep{2009ApJ...695..954L}; turbulent driving of the fractionation~\citep{2021FrASS...8....2R}; or multi-fluid effects in the solar chromosphere~\citep{Martínez-Sykora_2023}. The fractionation process is observed to operate on timescales of days, showing progression during the emergence and decay of active regions (ARs)~\citep[e.g.,][]{1992ESASP.348..347S,1995ApJ...440..884S,2001ApJ...555..426W}.

\begin{figure}[t]
    \centering
    \includegraphics[width=.805\linewidth]{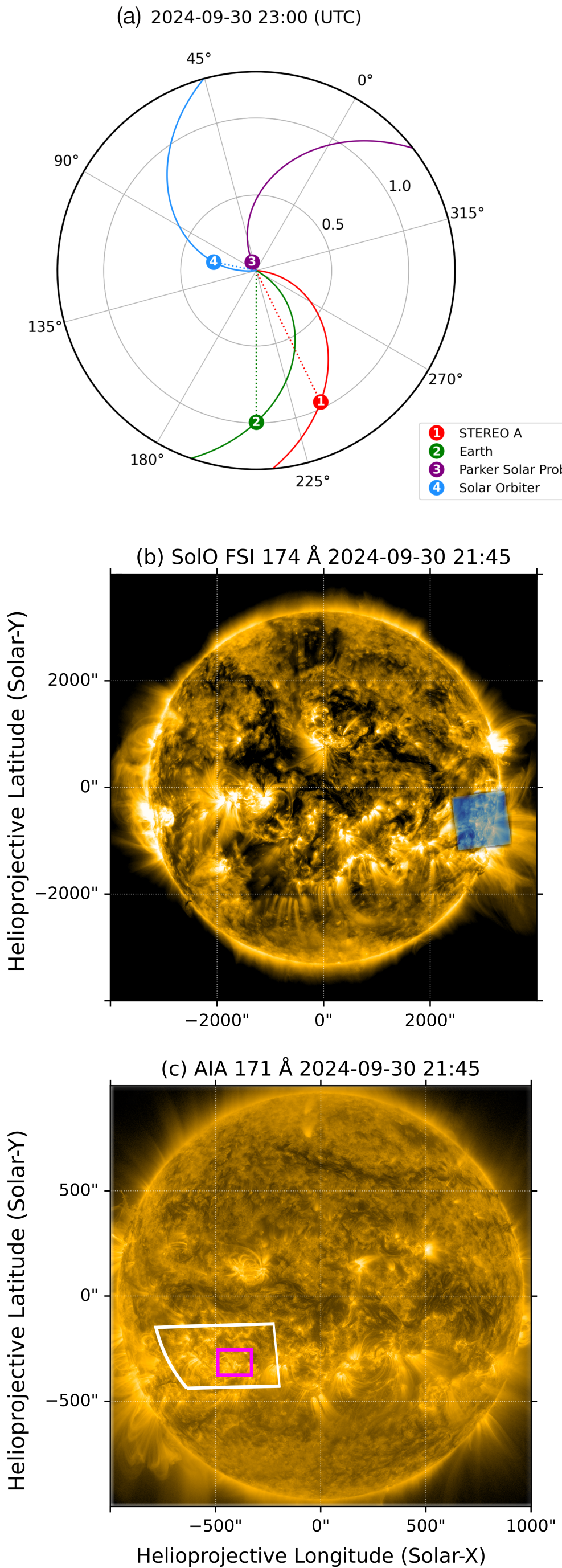}
    \caption{Vantage points of the observatories on 2024 Sept 30. Panel (a): Position of SO with
    respect to the Sun-Earth line (from Solar-MACH). Panel (b): The Sun observed with EUI/FSI on 2024 Sept 30 21:45\,UT with the FOV of EUI/HRI overlaid as seen at 23:05\,UT. Panel (c): SDO/AIA observation at 2024 Sept 30 21:45 UT,
    with EUI/HRI FOV overlaid as the white region and the Hinode/EIS FOV as the magenta one.}
    \label{fig:Overview_Observatory}
\end{figure}

Studies of solar flares, on the other hand, have reported that the plasma composition changes over timescales of minutes, with different parts of the coronal 
structures exhibiting different compositions during the flare~\citep{1981MNRAS.197...41V,1995ApJ...447..936F,2014ApJ...786L...2W}. However, there is an inconsistency between  measurements of the change in FIP bias from Sun-as-a-star X-ray observations versus spatially resolved EUV observations during flares. In particular,
the overview Table~1 in \citet{To_2024} illustrates that the 
differing conclusions of composition evolution during flares depend on the types of observations (X-ray or EUV). This is perhaps not surprising, because X-rays and EUV are preferentially sensitive to different temperature ranges \citep[e.g.,][]{winebarger2012}, and hot flare emission includes contributions both from directly-heated coronal plasma and from impact-heated chromospheric plasma that expands into the corona (``chromospheric evaporation'') \citep[e.g.,][]{fletcher2011, holman2011}. X-ray and EUV Sun-as-a-star flare observations show that elemental abundances evolve from coronal to near-photospheric values during the impulsive phase and rapidly return to coronal values in the decay phase for Mg, Al, Si, S, and Fe \citep{1984Natur.310..665S,2021ApJ...920....4M, 2014ApJ...786L...2W, 2023ApJ...957...14S}. However, the specific evolution of the observed FIP bias differs from element to element, and from flare to flare, and therefore the relative contributions of each process remain unclear~\citep{2015ApJ...803...67D}.
Even more interestingly,
during solar and stellar flares, an inverse FIP effect is sometimes observed during 
and after flaring activity \citep{Doschek_2015,2019ApJ...875...35B}, which could be a signature for the reversal of the fractionation process. The inverse FIP effect can be explained by the ponderomotive force model by \citet{2021ApJ...909...17L}, with different behaviors of the Alfv\'enic wave reflection in different magnetic field topologies. Due to the rapid evolution of the magnetic field and strong plasma flows during flares, the dynamics of the changing plasma composition during solar flares is still not well understood. In particular, the relative contribution of chromospheric evaporation flows and of the reconnection flows to the observed composition changes during solar flares, compared to other potential mechanisms, is still not well constrained~\citep{To_2024}.

In this article, we present Solar Orbiter \citep[SO;][]{Muller2020}
observations of an M7.6-class solar flare observed
on 2024 September 30, close to the western solar limb from the perspective of SO, comprising a suitable dataset for inferring plasma composition during flare progression. The Spectral Imaging of the Coronal Environment instrument \citep[SPICE;][]{2020A&A...642A..14S} and the Extreme Ultraviolet Imager \citep[EUI;][]{Rochus2020} instruments aboard SO captured the pre-eruptive, impulsive, and parts of the decay phases of the aforementioned flare. Due to the multi-vantage 
point observations from Earth and SO at quadrature, we can 
stereoscopically reconstruct the flaring AR and study 
the composition evolution before, during, and after the impulsive phase of the event. 
SPICE observed a range of spectral lines with temperature coverage in the range of 10$^5$--10$^6$\,K, suitable for differential emission measure (DEM) analysis ~\citep{Varesano_2024,Varesano_2025}. This flare is well observed with other observatories, including the Hinode Extreme Ultraviolet Imaging Spectrometer \citep[EIS;][]{2007SoPh..243...19C} instrument and the Solar X-ray Monitor (XSM) on the Chandrayaan-2~\citep{VADAWALE20142021, Mithun_2020SoPh..295..139M} lunar orbiter. To provide context for the integrated X-ray observations from XSM, we utilize co-temporal observations from the Spectrometer Telescope for Imaging X-rays \citep[STIX;][]{2020A&A...642A..15K} onboard Solar Orbiter, that observes the flaring region in quadrature as well.

In Section~\ref{sec:Observations}, we present the observations and the 
reduction of the data. In Section~\ref{sec:data_processing}, we describe the DEM 
processing and the novel technique for abundance analysis from spectral observations used in our 
workflow, as well as the magnetic reconstruction of the active region. In 
Section~\ref{sec:Abundances}, we describe the inferred abundance evolution and 
the corresponding results, with our conclusions outlined in
Section~\ref{sec:Conclusions}.

\section{Observations}
\label{sec:Observations}

On 2024 September~30 at 23:37\,UT, 
Solar Orbiter observed an M7.6-class solar flare associated
with NOAA AR\,13842 close to the western limb from the vantage of SO, and
close to the disk center from Earth's perspective. Figure~\ref{fig:Overview_Observatory}, top panel \citep[produced with Solar MACH;][]{Gieseler_SolarMACH_2022}, illustrates the favorable observing geometry, as the projection for the SO vantage point provided an almost sideways observation of the flare close to the limb
while also providing footpoint information. The 
region observed by the EUI aboard SO was also observed from Earth, denoted as the area bounded in white
in panel~(c) of Figure~\ref{fig:Overview_Observatory}, which shows data from the Solar Dynamics Observatory Atmospheric Imaging Assembly instrument~\citep[AIA;][]{2012SoPh..275...17L} in the corresponding 171\,{\AA} channel. 
Figure~\ref{fig:Overview_Observatory} shows the closest in time
exposure from the EUI Full-Sun Imager (EUI/FSI), showing the full Sun for context in panel (b).

\begin{figure}
    \centering
    \includegraphics[width=\linewidth]{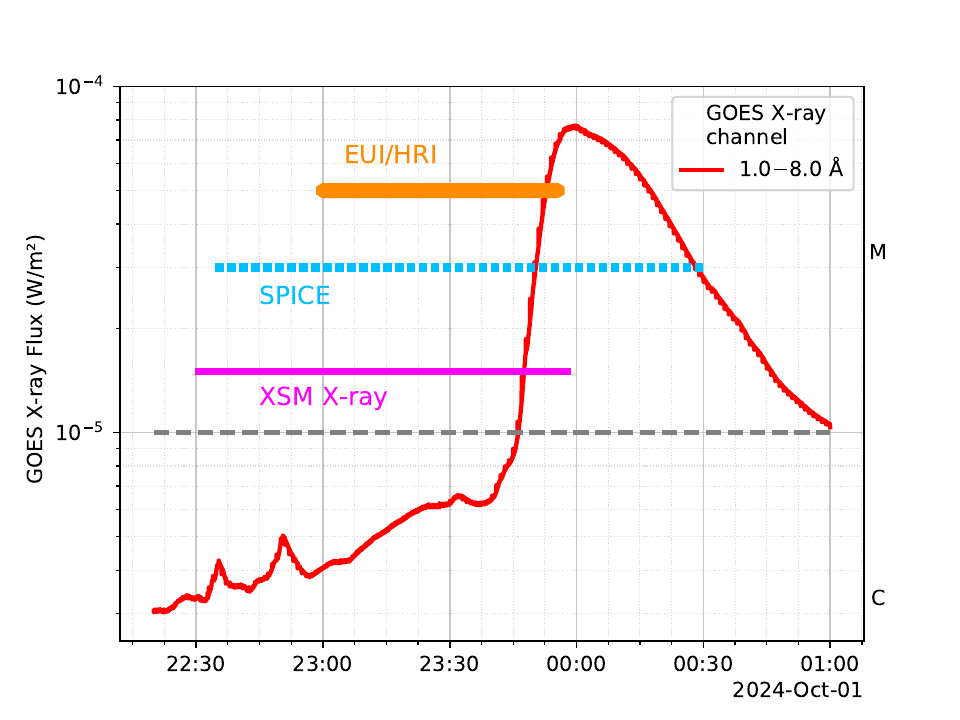}
    \caption{GOES X-ray light curve of the flare, where the right-hand vertical axis shows the flare classification. Overplotted are the exact times of observations from SO/SPICE and EUI instruments, as well as the Chandrayaan-2/XSM X-ray instrument observations used in this work.}
    \label{fig:X-ray_curve}
\end{figure}

\begin{figure}
    \centering
    \includegraphics[width=\linewidth]{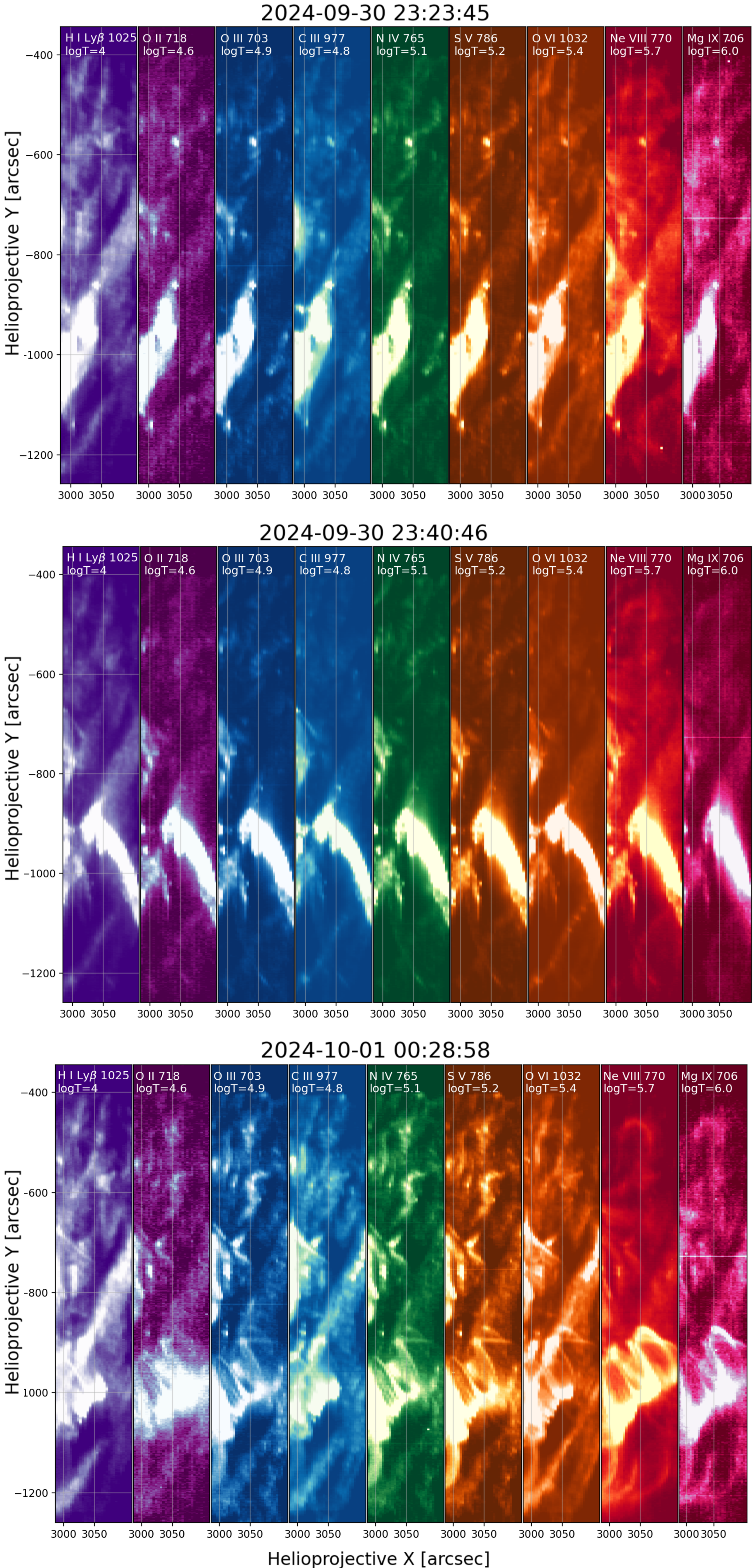}
    \caption{Overview of the observed spectral line intensities during the flare on 2024 Sept~30 with SPICE. The three time instances in panels (a), (b), and (c) correspond to the 
    pre-flare stage, impulsive phase including the acceleration of the ejecta, and post-flare loop contraction stage, respectively. The total intensity in each spectral line is shown, labeled by subtitles in each panel. A video of the observations showing the flare evolution is 
    available in the supplementary online materials.}
    \label{fig:SPICE-Overview}
\end{figure}

During this observing campaign, EUI/HRI 174\,{\AA} observed with a 2-second cadence during 2024 September~30 22:59--23:55\,UT (see Figure~\ref{fig:X-ray_curve} for the overlaid observation timelines). SPICE observed in a flare configuration with a fast raster strategy, with exposure of 20~s and 18-step raster resulting in 2~min 58~s cadence; the brightest spectral 
lines SPICE observed are shown in Figure~\ref{fig:SPICE-Overview}. SPICE was able
to observe the flare before its impulsive phase (Figure~\ref{fig:SPICE-Overview}, panel (a)), all the way
through the decay phase of the flare, as seen in panel (c) of Figure~\ref{fig:SPICE-Overview}.
EUI/HRI observed the pre-flare and impulsive phases of
the eruption, but stopped observing about 30~minutes before 
SPICE ended its observations, meaning that the last 18~SPICE rasters do not have context imaging from EUI. We aligned the SPICE data to the EUI observations with a cross-correlation approach using diagnostics with similar temperature sensitivity (\ion{Ne}{8} 770\,{\AA} SPICE data to the EUI 174\,{\AA} ones), finding very good accordance after applying a shift of approximately +40{\arcsec} in the helioprojective longitude of the SPICE observations. 

\section{Data processing and analysis}
\label{sec:data_processing}

To extract the physical properties of the flaring plasma requires careful data processing, which we describe in the following section. This step is essential due to the different types of observables comprising this study.

\subsection{Reducing the SPICE data}
\label{subsec:SPICE_data_reduction}

\begin{figure}
    \centering
    \includegraphics[width=\linewidth]{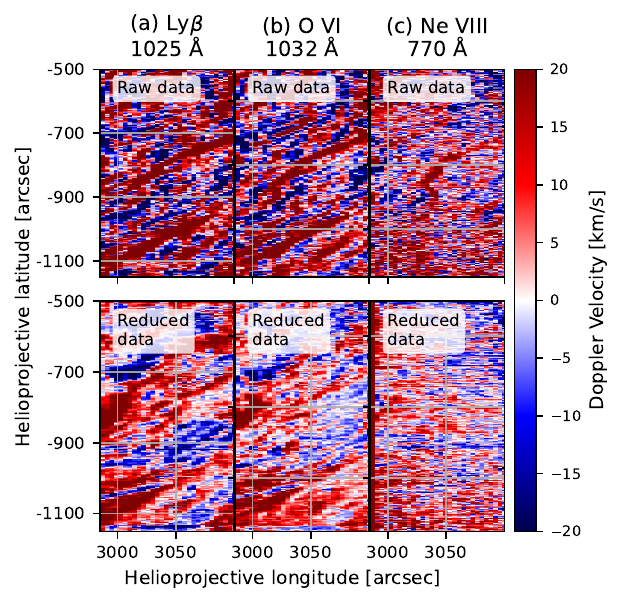}
    \caption{The SPICE PSF correction mitigates the spurious velocity signals in 
    the data. The top row shows the Doppler velocity derived from the uncorrected SPICE data, 
    whereas in the bottom row we show the resulting Doppler velocities after the 
    PSF artifact removal was applied \citep{Plowman_2025}.}
    \label{fig:PSF_removal_example}
\end{figure}

\begin{figure*}
    \centering
    \includegraphics[width=\linewidth]{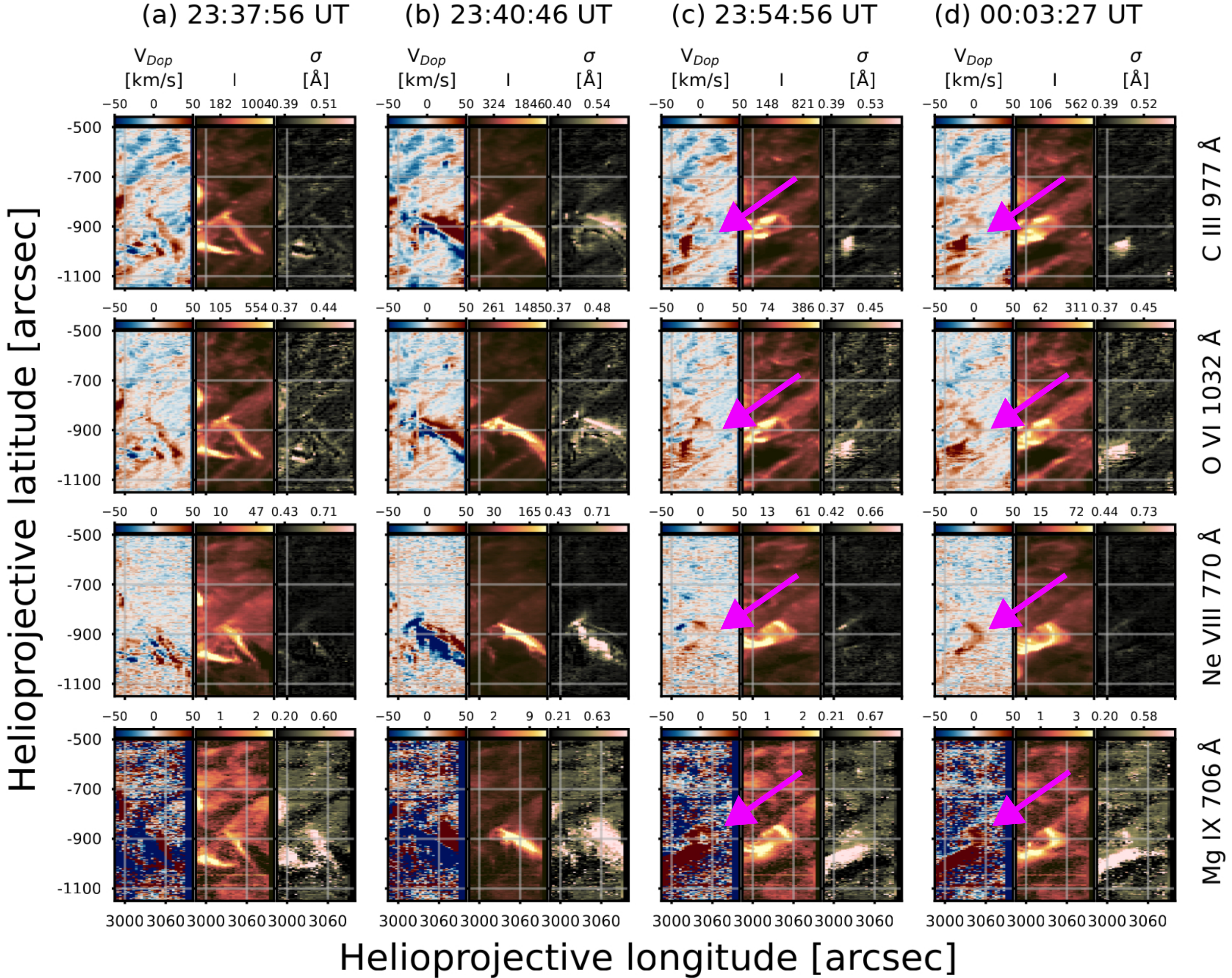}
    \caption{Overview of the spectral line
    parameters observed with SPICE during the flare. Each row represents a spectral line, noted on the right, where the panels show the temporal evolution with time to the right, as 
    noted on the panel titles. SPICE observations cover well the flaring AR during the impulsive (panels (a)-(c)) and gradual phases (panel (d)). The three columns in each subpanel correspond to Doppler velocity (in km/s), line peak intensity, and line width (in {\AA}). The magenta arrows show the region of interest in the Doppler maps, pointing to the signatures of reconnection downflows at the loop tops.}
    \label{fig:SPICE-Observation-overview}
\end{figure*}

SPICE is the EUV imaging spectrometer aboard Solar Orbiter, designed to obtain observations ranging from the upper chromosphere to the low corona in multiple lines of atomic species with both low and high FIP.
The SPICE data suffer from a non-uniform point-spread function (PSF) in the $x-y-\lambda$ space, requiring 
further reduction steps to alleviate this issue and make the data 
usable for further analysis. \citet{Plowman_2023} describe a way to remove the PSF through a minimization procedure, which requires a detailed knowledge of the instantaneous PSF
of the instrument. 
Based on this,
\citet{Plowman_2025} demonstrated that, through a deskewing of the apparent effect of
the PSF in the $x-y-\lambda$ space, one can remove the artifacts of the
PSF by estimating it through the data themselves. In this approach, the
method estimates a model PSF such that it minimizes the Doppler 
velocity across the whole FOV of the data. We have performed this step on a quiet-Sun 
dataset close to disk center that was taken on the same day, 2024 Sept 30 (as shown appropriate by 
\citealt{Plowman_2025}), where the expected average Doppler velocity should be close to 
zero. This estimate of the PSF was used for the reduction of the flare rasters close 
to the limb. Figure~\ref{fig:PSF_removal_example} shows results from the SPICE data 
reduction, for three of the spectral lines used in this study, selected from the two 
different detectors in the SPICE instrument. After applying the reduction, we found a 
shift of [--5, +1]\,pixels/{\AA}, making the rightmost two columns of the 
raster unusable. This result is consistent with the previously found reduction 
parameters in \citet{Plowman_2025} close to SO perihelia.
After the processing, as shown in the bottom panels of 
Figure~\ref{fig:PSF_removal_example}, 
the corrected data no longer show the prominent striping/streaking artifacts. 
We use the deskewed SPICE data as the basis for further analysis. Overviews  
of the physical parameters inferred from the SPICE data during the flare progression 
are shown in Figure~\ref{fig:SPICE-Observation-overview}, where the lines are ordered 
vertically in increasing temperature of formation.

\subsection{Hinode EIS analysis}
\label{subsec:Hinode_EIS_analysis}

Hinode/EIS is an EUV slit scanning spectrograph, observing a wide range of spectral lines with temperature sensitivity between 10$^4$ and 10$^7$\,K \citep{2007SoPh..243...19C}.
Furthermore, EIS observes many different lines of atomic species with 
different FIP biases, complementary to the current SPICE observations \citep{2024ApJ...976..188B}. Hinode/EIS observed during the flare and produced 
a raster scan over the AR as shown in Figure~\ref{fig:Overview_Observatory} as the magenta rectangle. 
 EIS scanned the AR right as the impulsive phase of the flare was initiating, with the spectral scan taking place during 2024 Sept~30 22:52--23:58\,UT, with the slit scanning from west to east (right to left) as discussed in Section~\ref{subsec:Abundance_evolution_X-rays_EIS}. The most recent EIS calibration from \citet{2023ApJS..265...11D} was used for the data reduction and the default \texttt{eis\_aia\_offsets} routine was applied for an  approximate alignment to AIA. The line ratio of \ion{S}{10} 264.3\,{\AA} and \ion{Si}{10} 258.4\,{\AA} is used as a FIP bias proxy derived from the Hinode data.


\subsection{Chandrayaan-2/XSM observations}
\label{subsec:XSM_analysis}

XSM is a soft X-ray instrument on board the Chandrayaan-2 mission to the Moon \citep{VADAWALE20142021,2020SoPh..295..139M}. XSM 
 observes the Sun-as-a-star and measures the solar X-ray spectrum in the energy range of 1--15\,keV with an energy resolution of 0.175\,keV (FWHM) at 5.9\,keV and time cadence of 1\,s \citep{Mithun_2021ExA....51...33M}. The broadband soft X-ray spectra from XSM provide diagnostics of the solar corona at various levels of solar activity from quiescent Sun to large solar flares.
 XSM observes spectral lines of low-FIP elements (Mg, Al, Si, Ca, and Fe), the mid-FIP element S, and the high-FIP element Ar, along with the continuum. This allows measurements of the time evolution of plasma temperature and elemental abundances. XSM observations have been used extensively to derive elemental abundances in quiet-Sun X-ray bright points \citep{Vadawale_2021ApJ...912L..12V}, ARs \citep{Zanna_2022ApJ...934..159D}, and solar flares \citep{2021ApJ...920....4M,  2025arXiv251002102M}.

XSM observed the Sun in the 1–15 keV energy range during the impulsive phase of the flare until 23:58\,UT, as noted in Figure~\ref{fig:X-ray_curve}. After that, as the flare flux increased further, an attenuator was inserted to reduce the photon flux. With the attenuator in place, XSM spectra were available only from 2\,keV onward. Therefore, in the present work, we analyzed XSM spectra only during the impulsive phase of the flare, when the full spectrum was available.
The raw XSM data were reduced with the \texttt{xsmdas} reduction package provided by the instrument team \citep{2021A&C....3400449M}. The resulting X-ray spectra 
were then analyzed with the \texttt{XSPEC} spectral fitting package~\citep{1996ASPC..101...17A}, which also provides the uncertainties of the inferred quantities quoted in this study. We used the \texttt{chisoth} model for fitting the X-ray spectra based on the ``astrophysical plasma with CHIANTI database'' approach described in \citet{2021ApJ...920....4M}. 
 We fitted the 1-minute cadence XSM spectra with a two-temperature component model, considering the temperature and emission measure of each component as free parameters, while the elemental abundances of Mg, Si, S, Ar, Ca, and Fe were treated as free parameters; the same abundances were applied to both thermal components.
These elements have prominent spectral lines in the observed XSM spectrum; the rest of the elements were assumed to have photospheric abundances \citep[taken from][]{2021A&A...653A.141A}.

\subsection{STIX observations}
\label{subsec:STIX_observations}

STIX is a hard X-ray imaging instrument on Solar Orbiter with an energy range of 4--150\,keV. In this study we use 
the data from the lowest energy bands (6--10\,keV) observed with STIX to localize the sources of the Sun-as-a-star signals detected with XSM. STIX observed during the entire flare and imaging reconstruction was available through the standard pipeline. A detailed description of the STIX observations and their reduction is given in \citet{2026A&A...705A.113C}.

\subsection{DEM and Abundance Analysis with SPICE}
\label{subsec:DEM_Ab_analysis}

\begin{figure*}[htp]
    \centering
    \includegraphics[width=\linewidth]{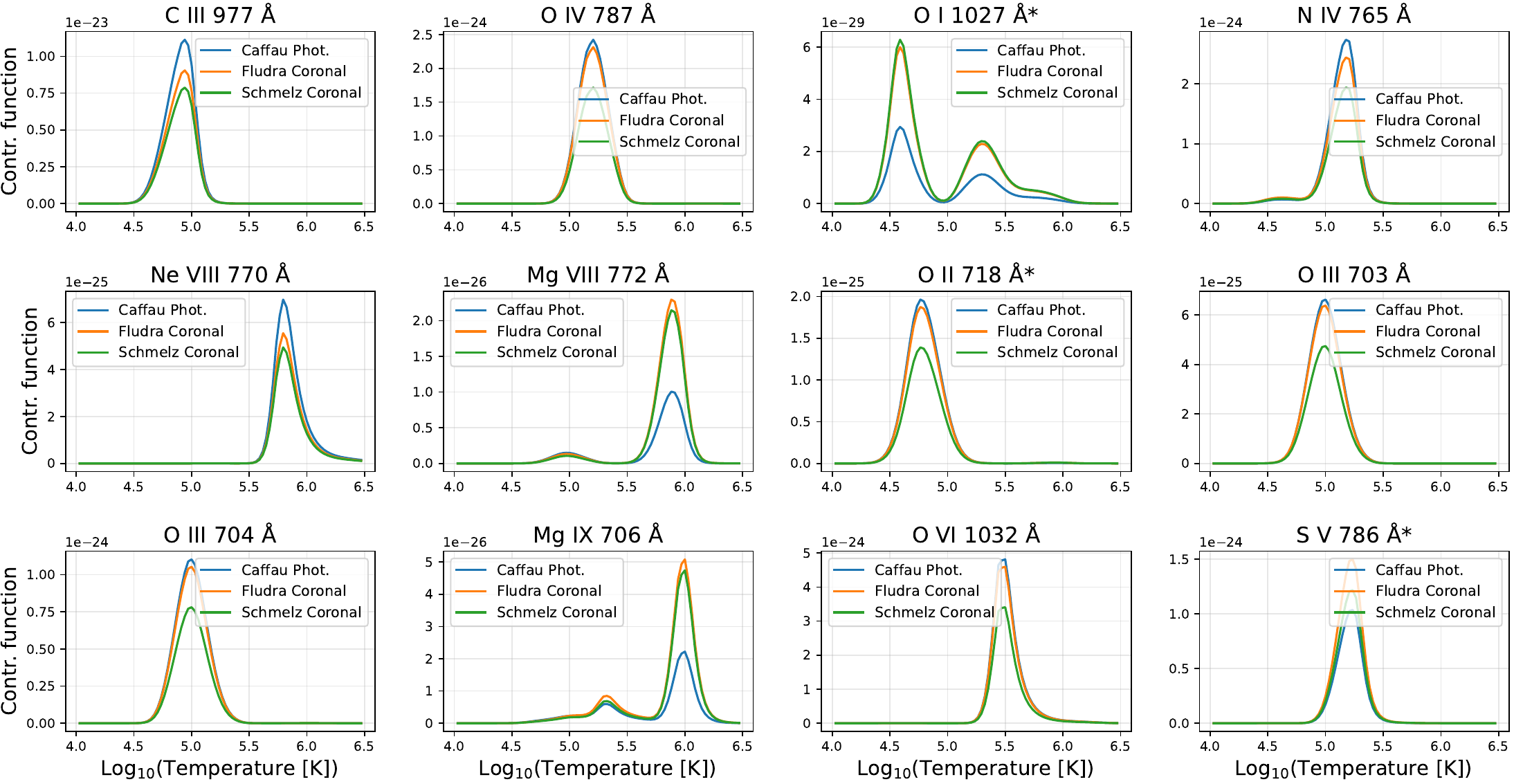}
    \caption{Temperature contribution functions from CHIANTI for the SPICE spectral lines used in 
    this study. The differently colored lines correspond to the different 
    solar abundance models adopted in the contribution function computations.
    The spectral lines with an asterisk in their title are not used in the combined DEM+abundance analysis, described in Section~\ref{subsec:DEM_Ab_analysis}.}
    \label{fig:SPICE_lines_resp_fn}
\end{figure*}

SPICE observes many optically thin spectral lines formed in the transition region and corona, which are listed in Figure~\ref{fig:SPICE_lines_resp_fn}. The overlapping temperature sensitivity of the SPICE lines allows for
DEM reconstruction between $T$ of 10$^5$--10$^6$\,K. We use the ``column'' DEM, which is the squared density of material that emits along the line of sight, integrated along the line of sight, in a given temperature range \citep{1976A&A....49..239C}. The assumption for applying DEM analysis is that the optically thin lines are collisionally excited, which generally holds for the low corona and transition region, where the SPICE lines are formed. Radiative excitation, which becomes significant at higher altitudes \citep{Seaton2025}, is assumed negligible here. We have omitted from our analysis the optically thick emission lines of \ion{H}{1}, \ion{O}{1}, and \ion{O}{2}, which are present in the SPICE data but form at lower temperatures, as noted in Figure~\ref{fig:SPICE_lines_resp_fn}.

We use the DEM algorithm of 
\citet{Plowman2020} via the \texttt{EMToolkit} interface \citep{Plowman_EMToolKit_A_Standardized} to reconstruct the DEM from the SPICE lines. The temperature contribution functions are precomputed for the SPICE lines, shown in Figure~\ref{fig:SPICE_lines_resp_fn} using CHIANTI~v8.0.7 \citep{Dere97, ChiantiPy} assuming electron density of 10$^{8}$\,cm$^{-3}$, where we have used the Python wrapper \texttt{fiasco} \citep{Barnes2024} to perform these computations.\footnote{Comparison of the contribution functions with CHIANTI v11 showed no significant difference with the presented results.}
The \ion{S}{5} line was omitted because we will utilize it as a cross-check of the robustness of our method and 
to study its behavior as an intermediate FIP element. We have also omitted it from the DEM calculation as we do not want it to affect the inferred DEM distribution, which will in turn affect its inferred abundance evolution, while other lines such as the \ion{O}{4} 787\,{\AA} and \ion{N}{4} 765\,{\AA} have similar contribution functions that constrain the DEM in this temperature range.

The coronal plasma abundance based on the SPICE spectral observations is derived from a DEM inversion approach, where the abundance is left as a free parameter. In this approach, the abundances of all elements vary together between the values of a photospheric~\citep{2008A&A...488.1031C} to coronal abundance models~\citep{1999A&A...348..286F}. The inversion technique varies a single free parameter that linearly interpolates between the two abundance models, and that parameter effectively determines all of the elemental abundances. 
For each pixel, in each timestep, we recorded which abundance value and DEM has the minimal $\chi^2$ as a solution for the given SPICE observations. After finding the abundance and corresponding DEM that best fit the observations through the analysis above, we examined the behavior
of sulfur during the flare through a 2-line ratio method accounting for the different contribution functions as outlined in Section~\ref{subsec:Abundance_evolution_S_Mg}.
The two-line approach takes into account the plasma temperature distribution, the atomic parameters of the 
transition (contribution functions), as well as the instrument response \citep[e.g.,][]{ZambranaPrado_2019, Varesano_2024}:  

\begin{equation}
\frac{A_{LF}}{A_{HF}} = \frac{I_{LF}}{I_{HF}} \left ( \frac{A_{LF}^{ph}}{A_{HF}^{ph}}\frac{<C_{LF}(T), DEM(T)>}{<C_{HF}(T), DEM(T)>} \right) ^{-1}
\end{equation}

where the FIP bias, taken as the ratio of the low-FIP (LF) to high-FIP (HF) abundances $\frac{A_{LF}}{A_{HF}}$, is proportional to the line intensities $I$, the photospheric abundances $A^{ph}$, and the convolution of the line contribution function (derived from CHIANTI) and the inferred DEM from Section~\ref{subsec:DEM_Ab_analysis} over the temperature range. We have used the DEM distribution from the best fit to the data (lowest $\chi^2$), while using the photospheric line contribution functions (see Figure~\ref{fig:SPICE_lines_resp_fn}). Synthetic recovery tests using the adopted SPICE line set show that the abundance parameter is well constrained over the central part of the instrument’s temperature-response range, while the recovery becomes less reliable near the temperature boundaries where fewer low- and high-FIP lines provide overlapping constraints.

\section{Coronal plasma abundance Evolution during the flare}
\label{sec:Abundances}

Combining observations from all instruments allows us to present a comprehensive 
picture of the flaring plasma evolution. In this section, we describe how the combined 
observations provide insights into the elemental abundance changes of different regions of the flaring AR. 

\subsection{Abundance evolution from the SPICE line analysis}
\label{subsec:Abundance_evolution_all_lines}

\begin{figure*}
        \includegraphics[width=\linewidth]{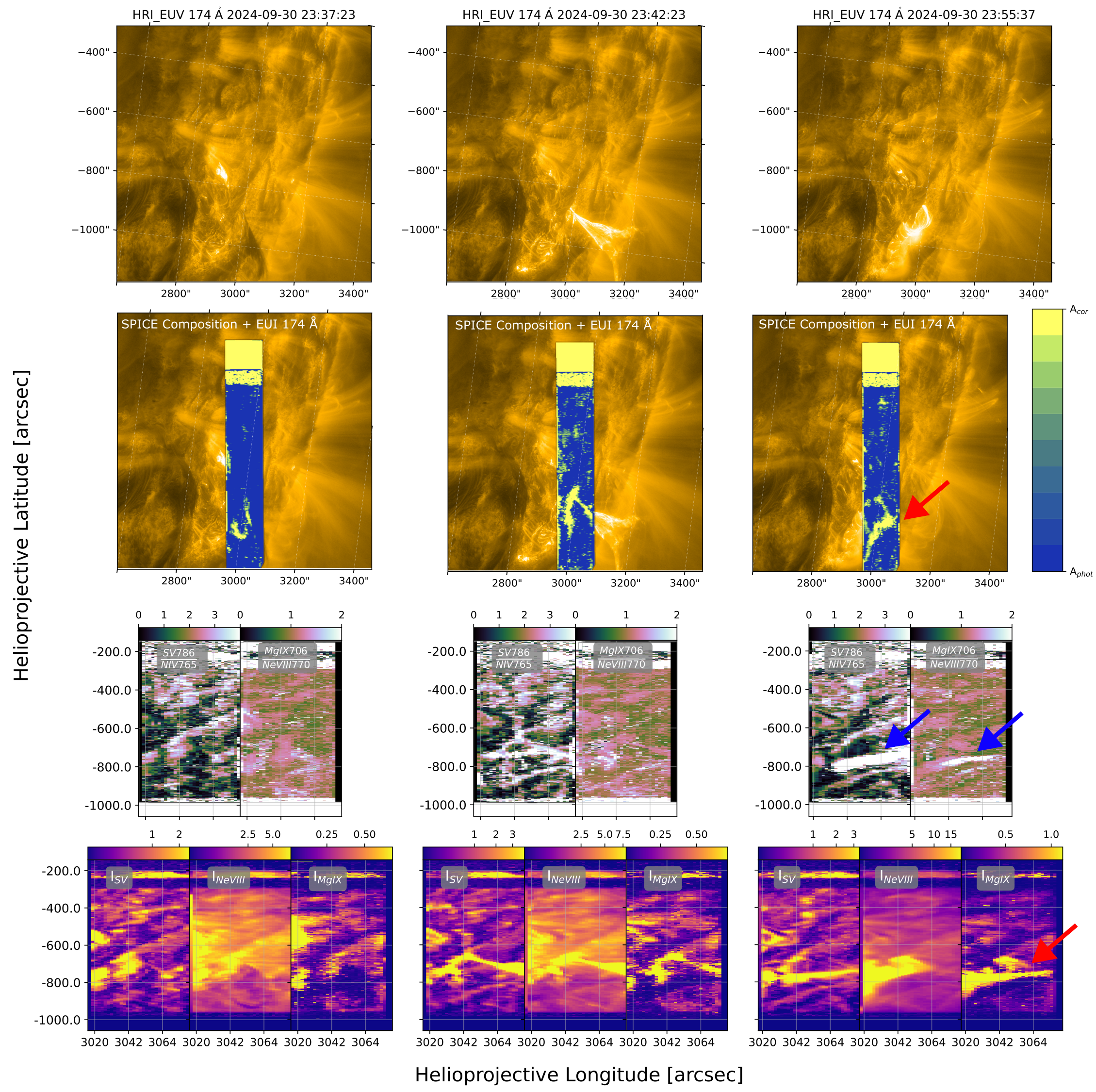}
    \caption{Evolution of the coronal plasma composition during the flare, for the pre-eruption (left column), impulsive phase (middle column), and gradual phase (right column). Each column corresponds to the same time instance. \emph{First row}: EUI/HRI 174\,{\AA} observations of the flare progression. \emph{Second row:} The SPICE inferred plasma abundance overlaid onto the EUI images, with blue indicating photospheric composition and yellow indicating coronal composition. \emph{Third row}: The time evolution of FIP-bias derived from the 2-line method, for the line ratios \ion{S}{5} 786\,{\AA} to \ion{N}{4} 765\,{\AA}, and \ion{Mg}{9} 706\,{\AA} to \ion{Ne}{8} 770\,{\AA}, for the same three time instances. \emph{Fourth row}: Peak intensity of the SPICE lines \ion{S}{5} 786\,{\AA}, \ion{Ne}{8} 770\,{\AA}, \ion{Mg}{9} 706\,{\AA} at the corresponding times.}
    \label{fig:Abundance_SO}
\end{figure*}

Based on the DEM+abundance inversion method described in Section~\ref{subsec:DEM_Ab_analysis}, we computed the varying plasma abundance evolution in the SPICE observations during the flare, which is shown in the second row of Figure~\ref{fig:Abundance_SO}. Bluish tones correspond to photospheric abundances, and yellow to coronal abundances. Note that our method infers that most of the FOV has photospheric abundance,  while the flaring region looptops exhibit coronal (low) FIP bias. This is corroborated by the Hinode/EIS observations, described in Section~\ref{subsec:Abundance_evolution_X-rays_EIS} and Figure~\ref{fig:Hinode_observation_analysis}. We find that the looptops become filled with plasma exhibiting increased FIP bias(noted as the red arrow on the second row in Figure~\ref{fig:Abundance_SO}), similar to the results shown in \citet{To_2024}. These changes occur on the order of a few scans of the SPICE raster, 
corresponding to about ten minutes. A caveat to be noted is that the observed EUV coronal loops must contain plasma at temperatures within the formation ranges of the spectral lines. Hence, the apparent evolution of the SPICE looptops might correspond to different loops coming into and out of the cooler temperature bands, where SPICE and EUI can detect them. The observations are consistent with downflows transporting coronal-composition plasma toward the tops of cooling post-flare loops as suggested in \citet{To_2024}. The origin of this downflow could be due to the reconnection outflow or coronal condensation; and signatures of it present in the SPICE Doppler velocities are further discussed in the following sections. During the progression of the flare and the abundance changes, strong 
blueshifts are present in the loop footpoints and redshifts at the tops of the loops. However, associating the observed Doppler velocity in the SPICE data with upflows and downflows is not straightforward due to the optically thin nature of the plasma and the projection of the complex geometry of the AR towards the observer. Hence, we address this issue with 3D modeling of the AR with the CROBAR method~\citep[see][for a detailed description]{CROBAR_2021} to estimate the inclination of the coronal loops to the observer in Section~\ref{subsec:CROBAR_results} and interpret the observed Doppler motions as upflows or downflows.

\subsection{Abundance evolution of the S and Mg composition during the flare}
\label{subsec:Abundance_evolution_S_Mg}

We have performed a separate analysis of the evolution of the FIP-sensitive line ratios \ion{S}{5} 786\,{\AA} to \ion{N}{4} 765\,{\AA} and \ion{Mg}{9} 706\,{\AA} to \ion{Ne}{8} 770\,{\AA}, to investigate the plasma composition evolution during the flare progression~\citep{Varesano_2025}. This estimate is independent from the one presented in Section~\ref{subsec:Abundance_evolution_all_lines}, where we have utilized the resulting DEM from the analysis in Section~\ref{subsec:DEM_Ab_analysis} only. We have utilized the two-line approach, that was described previously in Section~\ref{subsec:DEM_Ab_analysis}.

The resulting FIP biases, presented in the third row of Figure~\ref{fig:Abundance_SO}, show clearly the abundance evolution during the flare progression. The lines used in these ratios form at lower temperature than those usually used in the Hinode/EIS EUV analysis and are therefore a complementary diagnostic \citep{2024ApJ...976..188B}. The principal result from these observations is that the looptops (highlighted with the blue arrows on the third row of Figure~\ref{fig:Abundance_SO}) get filled with increased FIP-bias plasma, corroborating the findings in Section~\ref{subsec:Abundance_evolution_all_lines}. This behavior is observed in both the cooler line ratio of the \ion{S}{5} 786\,{\AA} to \ion{N}{4} 765\,{\AA} as well as in the hotter \ion{Mg}{9} 706\,{\AA} to \ion{Ne}{8} 770\,{\AA}. In particular, this result shows the robustness of the DEM+abundance inversion method presented in Section~\ref{subsec:Abundance_evolution_all_lines}, as it retrieves the same abundance evolution conclusion. 

\subsection{Modeling of the AR structure with CROBAR}
\label{subsec:CROBAR_results}

\begin{figure*}
    \centering
    \includegraphics[width=\linewidth]{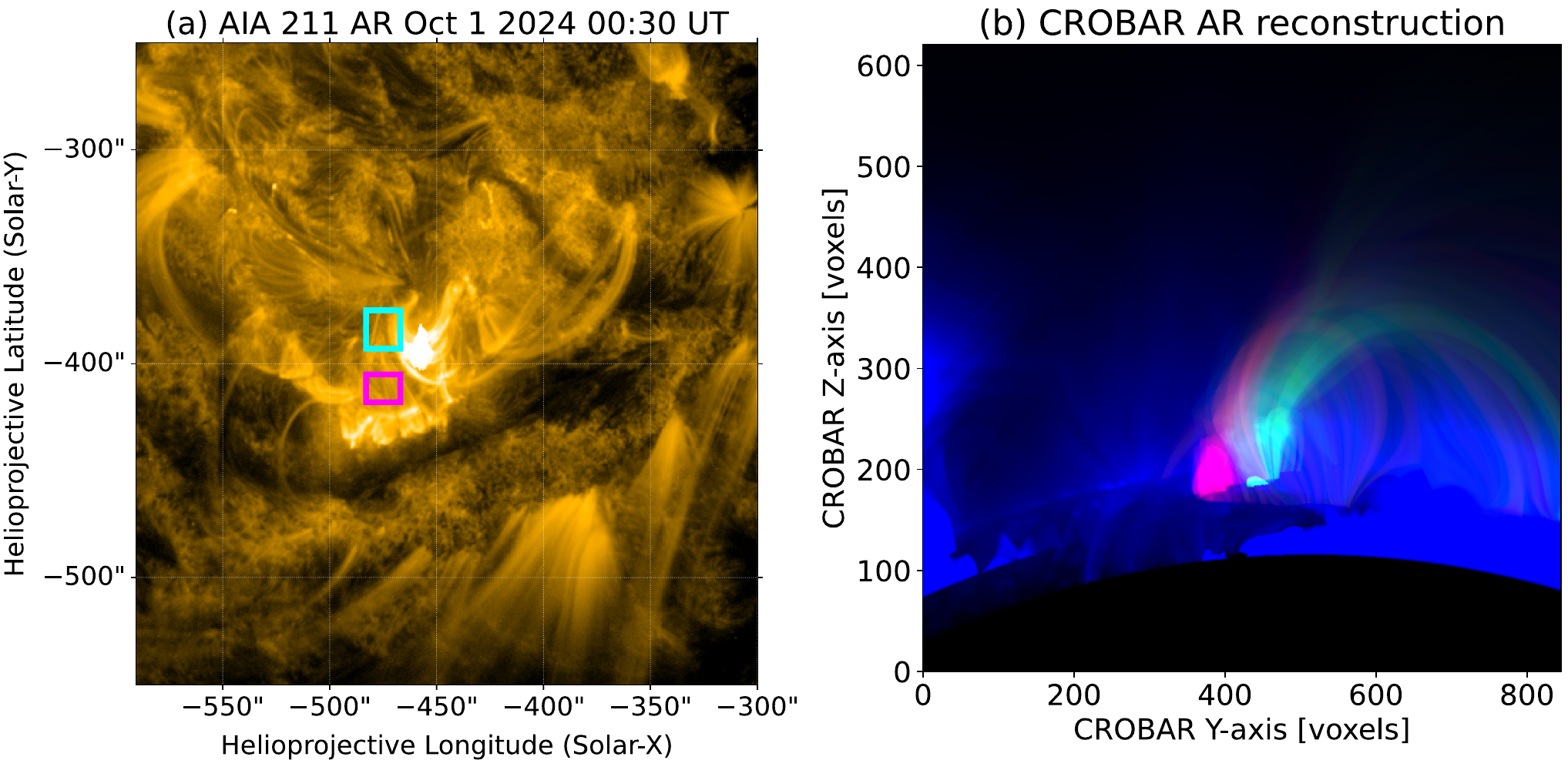}
    \caption{3D Reconstruction of NOAA AR~13842 with the CROBAR modeling framework. Panel (a): SDO AIA 211\,{\AA} cutout image of the modeled region. The magenta (looptops) and cyan (footpoints) rectangles show the regions of the FOV used for the field tracing. Panel (b): Reprojected CROBAR model of the AR as seen from above the Sun (North Pole orientation), where the illuminated field lines are connected to the flare footpoints in the reconstruction. Note the magenta ones correspond to the compact post-flare loops seen in SPICE.}
    \label{fig:CROBAR-Reconstruction}
\end{figure*}

To interpret the observed Doppler shifts in the SPICE data 
in terms of downflows and upflows, we use the CROBAR \citep{CROBAR_2021,CROBAR_2023} framework to construct a model of the AR based on magnetograms and EUV observations. 
CROBAR reconstructs the coronal magnetic field through a linear force-free field (LFFF) extrapolation of the observed
AR from a photospheric magnetogram, where in this case we used the available SDO/HMI line-of-sight magnetic field strength \citep{HMI_2012}. In this case, an LFFF approach of the post-flare loops is justified, as the highly twisted prominence has already erupted and the AR is in a stable state with lower helicity.
Then, CROBAR populates the volume with hot plasma along a volume-filling basis of coronal loops 
based on the magnetic field extrapolation to minimize the residual between
the observed coronal emission intensity (in this case AIA 211\,{\AA}) 
and the intensity predicted from the plasma-filled magnetic AR model. The resulting reconstruction provides
us with a 3D rendering of the AR, where each coronal loop has a known set of three-dimensional coordinates associated with it, as well as thermodynamic and magnetic plasma properties along the loop length~\citep{2026ApJ...997..293P}. The coronal loop reconstruction allows for reprojecting the model from the Earth-Sun line to the vantage point of SO and computing the apparent geometric orientation of the emitting coronal loops towards SPICE.

For this reconstruction, we have chosen the HMI magnetogram and AIA 211\,{\AA} image from 2024 Oct 1 00:30\,UT centered on the AR as seen from Earth, at helioprojective coordinates [--350{\arcsec}, --350{\arcsec}] with size of 500{\arcsec} $\times$ 500{\arcsec}. We have traced 7500~field lines across the domain, where the LFFF extrapolation uses a constant value of --7.5 turns per Gm, as this value best satisfies the observed AR geometry, with minimal $\chi^2$ metric. The results from this process are shown in Figure~\ref{fig:CROBAR-Reconstruction}.
To study the geometry of the particular post-flaring loops seen in Figure~\ref{fig:Abundance_SO}, we retrieved the magnetic field lines originating from the footpoints of the AR post-flare loops as well as the top of the flare arcade, as shown in panel~(a) of Figure~\ref{fig:CROBAR-Reconstruction} as the magenta and cyan squares respectively. The FOV of Figure~\ref{fig:CROBAR-Reconstruction} panel~(a) is a cropped subfield of the whole image used for the reconstruction. 
Then, we have traced these loops through the domain of the CROBAR reconstruction -- an illustration of these loops as seen from the solar North Pole (top view) is shown in Figure~\ref{fig:CROBAR-Reconstruction} panel (b). To estimate the orientation of the post-flare loops relative to the observer, Solar Orbiter in this case, we compute the dot-product of the orientation of each voxel of the coronal loop model in CROBAR and the unit vector (orientation) of Solar Orbiter. The Solar Orbiter ephemeris was retrieved from the JPL NAIF SPICE kernels \citep{ACTON199665, ACTON20189}.
These CROBAR results point us to the fact that footpoints of the post-flare loops closer to disk center seen by Solar Orbiter are oriented along the line-of-sight, and the ones closer to the limb are pointing away from Solar Orbiter. Hence, we can interpret the data of the blueshifts at the loop footpoints as chromospheric evaporation in the closer footpoints, noted in Figure~\ref{fig:SPICE-Observation-overview} as the magenta arrows. We do not discuss the farther away footpoints due to superposition of the coronal structures on top of them. On the other hand, the looptops are oriented along the LOS; hence, the redshifts observed above the looptops correspond to downflows. This result is consistent with the reconnection-driven outflow mass loading or coronal condensation above the looptops observed with SPICE. We favor the former explanation, as there is no detection of coronal condensantion above the coronal loop tops seen in the cooler SPICE lines, up to the \ion{Mg}{9} 706\,{\AA} line sensitive to plasma up to 1\,MK. 


\subsection{Comparison with abundance analysis from Hinode/EIS and Chandrayaan/XSM}
\label{subsec:Abundance_evolution_X-rays_EIS}

To compare the FIP-bias evolution observed in X-ray and EUV, which sample different regions of the flaring AR, we compare the SPICE observations with EUV ones from Hinode/EIS and  Chandrayaan-2/XSM X-ray data for the same event.

\begin{figure*}
\includegraphics[width=\linewidth]{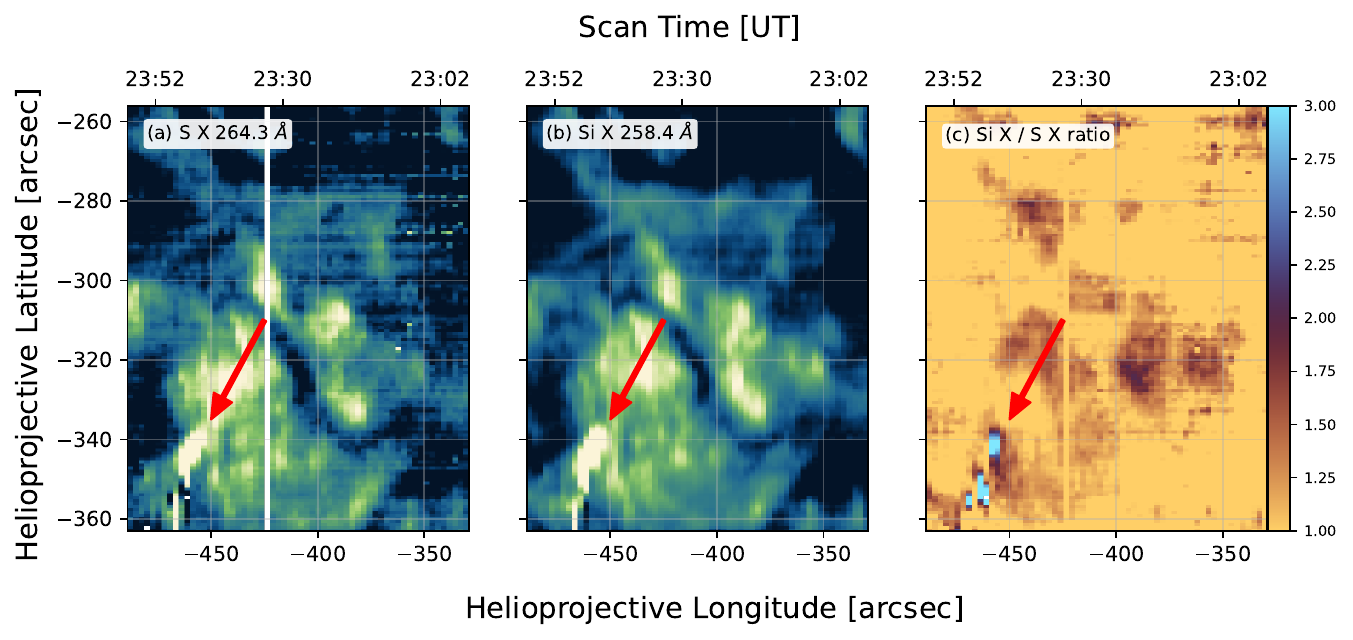}
    \caption{Hinode/EIS observations of the erupting AR in a pair of spectral lines with close formation temperatures, with low and high FIP, respectively. These observations were taken before, during, and after the flare, where the corresponding scanning time of the raster is noted in the top of the panels, exhibiting
    the west to east scanning direction of Hinode/EIS. Panel (a): \ion{S}{10} 264.3\,{\AA} line intensity; (b) \ion{Si}{10} 258.4\,{\AA} line intensity; (c) \ion{Si}{10} 258.4\,{\AA} to \ion{S}{10} 264.3\,{\AA} FIP-bias proxy. The red arrow highlights the location of the solar flare.}
    \label{fig:Hinode_observation_analysis}
\end{figure*}

Hinode/EIS observed the AR with a dense spectral raster during the impulsive phase of the flare. The observations of the intensities and the ratio of the FIP-bias-sensitive spectral line pair of \ion{Si}{10} 258.4\,{\AA} and \ion{S}{10} 264.3\,{\AA} are shown in Figure~\ref{fig:Hinode_observation_analysis}. The corresponding time for the slit location is noted on the top of each row. The left and central panels show the line intensity of the high- and low-FIP ions, respectively. The right panel shows the FIP-bias proxy derived from the ratio of the total line intensity of the two spectral lines, adjusted for their photospheric abundances and contribution functions~\citep{2008A&A...488.1031C}.
 The lines of \ion{S}{10} 264.3\,{\AA} and \ion{Si}{10} 258.4\,{\AA} have approximate line formation temperature of about $\textrm{log}_{10}(T/K)=6.17$ \citep{2023ApJ...959...72M}, hence we did not correct for the temperature dependence of the DEM distribution. The EIS lines provide a higher-temperature counterpart to the temperature range probed by the SPICE data \citep{2024ApJ...976..188B}. The EIS AR observations show an increase in intensity, which corresponds to the bright spot in the left side of the figure highlighted by the red arrow, at times after 23:30\,UT, when the flare enters its impulsive phase. The spectral fits across the raster are robust where the intensity signal-to-noise is high. We also investigated the line pair of high/low FIP elements \ion{Ar}{14} 194.40 \AA\  and  \ion{Ca}{14} 193.87\,{\AA}, but we found a weak signal-to-noise of the \ion{Ar}{14} 194.40 {\AA} line across the FOV, coupled with blending of a \ion{Fe}{10} line in the flaring region. These factors made the interpretation of this second FIP-bias sensitive line pair unreliable and is not further discussed. The \ion{Si}{10} 258.4\,{\AA} to \ion{S}{10} 264.3\,{\AA} ratio clearly shows the enhanced FIP bias in the flaring region, pointed out with the red arrow. This somewhat agrees with the SPICE conclusions, but the lower spatial and temporal resolution of EIS makes further comparison with SPICE challenging.

\begin{figure}
    \includegraphics[width=\linewidth,keepaspectratio]{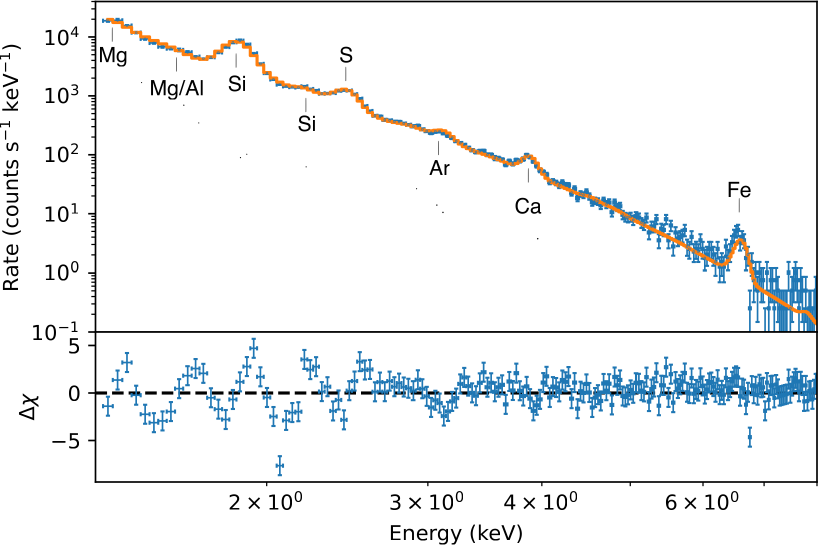}
    \caption{Example XSM spectrum (blue) for time 23:52\,UT fitted with two-temperature models from the \texttt{chisoth} package for the \texttt{XSPEC} spectral fitting software, where the contributing atomic spectral features are noted. The normalized difference between the model fit and the spectrum is plotted in the bottom subpanel.}
    \label{fig:XSM_example}
\end{figure}

\begin{figure}[h]
    \centering
    \includegraphics[width=\linewidth]{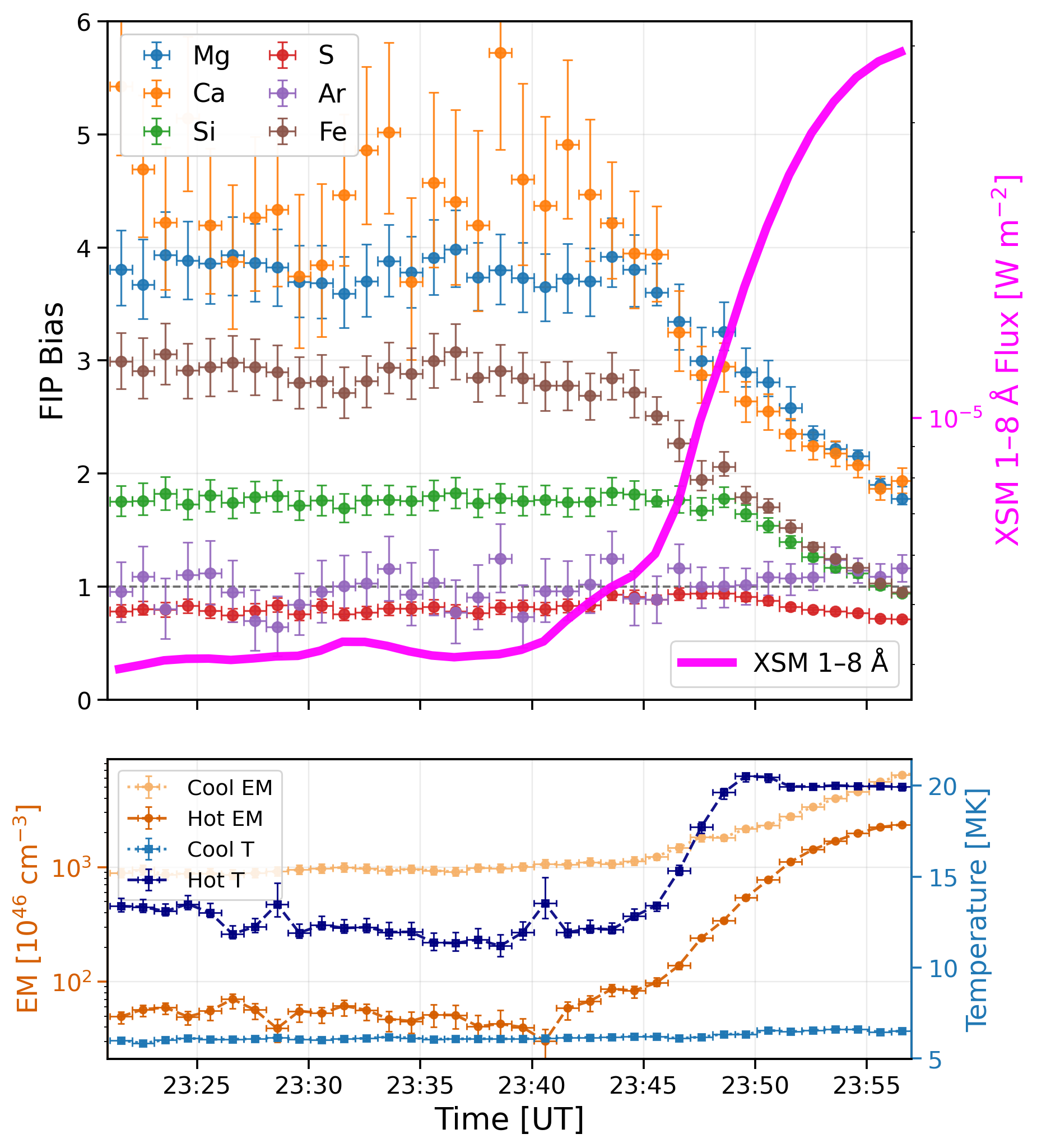}
    \caption{Top panel: Inferred FIP bias (the ratio of the inferred abundance over photospheric abundance model) from the X-ray XSM data for the different elements listed in the legend for the impulsive duration of the flare, for which XSM data is available (magenta curve). The low-FIP elements exhibit abundance decrease during the impulsive phase of the flare, whereas the high-FIP (Ar) element does not show any abundance evolution. Bottom panel: EM and temperature of the two components (cool and hot) of the fitted model.}
    \label{fig:XSM_abundances}
\end{figure}

To compare with X-ray observations of the same flare, we utilized the XSM data during the impulsive phase of the flare. XSM observes the solar soft X-ray spectrum with  disk-integrated measurements without spatial information. Furthermore, the XSM instrument utilizes an attenuator around the flare peaks, to avoid saturation of the detector. 
However, the attenuator removes lower-energy photons from the spectrum, and therefore abundance analysis during the peak of the flare is not possible, starting from 2024 Sept 30 23:58 UT onwards in the case of this event.

An example of the XSM spectrum is shown in Figure~\ref{fig:XSM_example}. In the top panel we show the spectral data for 1-minute integration, with the spectral features noted with their corresponding ions. The fitting is performed with the \texttt{XSPEC} software (orange line) and the residual normalized by the uncertainty between the fit and model is shown in the bottom panel. 
The abundance was inferred for 1-minute integration windows during 2024 September 30 23:22--23:58\,UT, and the results are shown in Figure~\ref{fig:XSM_abundances}, with the corresponding XSM soft X-ray lightcurve. 
The one-sigma uncertainties in the fitted parameters are computed using the standard \texttt{XSPEC} routines while fitting the data with a two-isothermal-component model. In the bottom panel we also show the changes in the EM and temperature of the two components of the model. 
During the impulsive phase of the flare we see a decrease in the abundances of all low-FIP elements, such as Mg, Si, Ca, and Fe, which is consistent with previous XSM observations (e.g.,~\citealp{2021ApJ...920....4M, 2022ApJ...939..112M, 2025arXiv251002102M}). Near the peak of the flare, Si, S, and Fe all show a decreasing abundance, which is consistent with previously reported observations of Fe abundances from XSM and Hinode/EIS for the same event~\citep{2026arXiv260504223Y}, where the authors measured the absolute Fe abundance using EIS line-to-continuum measurements.
Sulfur shows a more modest decrease compared to the other low-FIP elements. Argon, which is the only high-FIP element in our XSM analysis, does not show a statistically significant change during the impulsive phase of the flare. The behavior of Argon during flares has not been previously reported with XSM. The increase of the hot component EM in our model (see bottom panel in  Figure~\ref{fig:XSM_abundances}) is consistent with chromospheric evaporation material being heated, which has photospheric abundance. This newly heated material emitting in X-ray produces the drop in the FIP bias observed with XSM in Figure~\ref{fig:XSM_abundances} \citep{2025arXiv250814866M}. Previous observations of other flares have shown that after the peak of the flare, not reported here, the soft X-ray FIP-bias goes back to coronal level, which is due to the chromospheric material (having photospheric abundances) cooling down and only the cooler component of the quiescent coronal (increased FIP bias) plasma remaining \citep{1984Natur.310..665S, 2020SoPh..295..175N}.

\begin{figure*}
    \centering
\includegraphics[width=\linewidth]{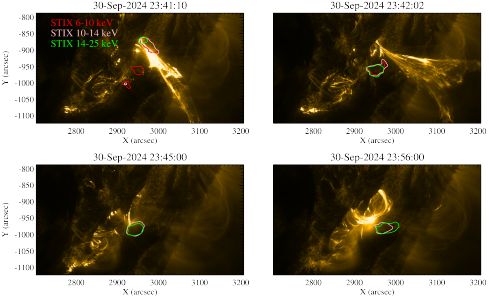}
    \caption{Evolution of the STIX X-ray sources overlaid on top of the EUI/HRI 174\,{\AA} images for four timesteps during the impulsive phase of the flare. The corresponding X-ray energy bins are outlined as the different colored contours, outlining the 60\% emission regions. We note that during the progression of the flare the location of the coronal X-ray sources corresponding to the XSM energies do not coincide with the EUV emission location. }
    \label{fig:STIX_data}
\end{figure*}

These XSM observations show differing abundance evolution from the SPICE observations, where the brightest parts of the post-flare loops show enhanced low-FIP abundances, as presented in Figure~\ref{fig:Abundance_SO}. To explain this difference, we use the imaging data from the imaging X-ray telescope STIX aboard Solar Orbiter to locate the source of the solar X-rays. The imaging STIX X-ray data for this flare, described in more detail in \citet{2026A&A...705A.113C}, are shown in Figure~\ref{fig:STIX_data}. We have overlaid the STIX 60\% intensity contours on top of the EUI 171\,{\AA} data, where the bright EUV loops correspond closely to the bright loops seen with SPICE. STIX is sensitive to more energetic X-rays (between 6 and 100\,keV) than XSM; in this study we use the lowest energy channels (6--10\,keV and 10--14\,keV) of STIX to compare with XSM, since this energy range is seen by both instruments. The STIX data show the source of the X-ray emission is located in a persistent location at solar helioprojective coordinates of about [+3000{\arcsec}, --980{\arcsec}] with a slight drift to the west, seen from Solar Orbiter. The X-ray radiation detected by XSM, is emitted from a source distinct from the bright EUV features. The X-ray STIX spectra, shown in Appendix A of ~\citet{2026A&A...705A.113C}, indicate that the 6--10\,keV energy band is dominated by thermal plasma emission. This emission is therefore likely coming from the plasma upflows associated with chromospheric evaporation, where photospheric FIP bias plasma is being ablated and elevated into the loops~\citep{2014ApJ...786L...2W,2025arXiv251002102M}. 
Hence, we can conclude that the discrepancy between the EUV and X-ray inferred abundance evolution is due to the different solar regions emitting the two types of solar radiation. In particular, XSM shows the dropping FIP bias consistent with chromospheric evaporation in the hottest loops, whereas SPICE is looking in the older, cooling, post-flare loops, which experience mass loading from above, as suggested from the detected SPICE Doppler shifts.

\section{Conclusions}
\label{sec:Conclusions}

We have described a set of observations of the M7.6-class 2024 Sept 30 flare with Solar Orbiter and a few other observing assets in the vicinity of Earth. The SO observations include SPICE data with a variety of spectral lines, suitable for analysis of the plasma composition during the flare 
evolution. 

After appropriate calibration of different datasets, we applied a combined approach to infer the DEM and composition simultaneously of the flaring plasma from the SPICE data. The approach, inverts at the same time for both DEM distribution and elemental abundance. The combined DEM+abundance inversion has the plasma abundance as a free parameter, linearly varying the composition of all elements between photospheric and coronal abundance model values. We found that the bright looptops seen by SPICE exhibit coronal FIP bias compared to the footpoints. When we computed the FIP bias for the \ion{Mg}{9} 706\,{\AA} to \ion{Ne}{8} 770\,{\AA}, and \ion{S}{5} 786\,{\AA} to \ion{N}{4} 765\,{\AA} lines using two-line ratios, we found similar results, corroborating the previous analysis. This distinct computation supports the robustness of the previous result 
based on the combined DEM+abundance inversion approach. Furthermore, using CROBAR we were able to reconstruct the geometry of the AR and associate the Doppler velocity observed in the SPICE data with the evolution of the plasma composition. The observed redshifts above the looptops correspond to downflows based on the CROBAR reconstruction, and are compatible with reconnection-driven downflows, which would transport plasma with coronal (enhanced) FIP bias. When these cooling looptops become visible in the SPICE spectral lines, they exhibit enhanced looptop FIP-bias, possibly due to the coronal material transport from the reconnection outflows. The presence of the plasma downflows is inferred from the redshifts present in the SPICE Doppler shifts, interpreted through the CROBAR AR modeling.

Comparison of the SPICE observations with other EUV and X-ray observations provides a complementary picture of the physical evolution of the plasma with different temperatures. The Hinode/EIS EUV observations, which only caught a very small part of the impulsive phase of the flare, also show an increase in the FIP bias in the flaring AR apparent in the \ion{Si}{10} 258.4\,{\AA} to \ion{S}{10} 264.3\,{\AA} ratio map in Figure~\ref{fig:Hinode_observation_analysis}. This result agrees with the finding from SPICE, that the brightest parts of the flaring regions exhibit increased FIP bias, despite measuring different temperatures. When comparing the flaring plasma composition detected with the Chandrayaan-2/XSM instrument, however, we find a significant discrepancy between the abundance evolution detected in the soft X-ray data with the SPICE EUV data. In particular, we find a decrease in the abundance of the low-FIP elements, and no change in the 
high-FIP element Ar abundance, as shown in Figure~\ref{fig:XSM_abundances}. To locate the X-ray sources of the XSM emission, we use SO/STIX imaging data. The location of the soft X-rays is displaced from the EUV loop tops as shown in Figure~\ref{fig:STIX_data}. Our results support the interpretation that the apparent EUV/X-ray discrepancy can arise because the diagnostics sample distinct thermal and spatial plasma components, possibly influenced by different processes. In particular, the X-ray data samples chromospheric material (with photospheric abundance) being evaporated into the corona and being heated, as suggested by the modeling of the XSM X-ray data presented in Figure~\ref{fig:XSM_abundances}. On the other hand, the SPICE EUV results show that the cooling post flare loops exhibit enhanced looptop FIP-bias. The observed downflows towards the looptops (as interpreted with the CROBAR modeling) are compatible with a reconnection downflow transporting coronal material (elevated  FIP-bias) to these looptops. These old loops were previously seen in the XSM data as the hot loops with photospheric FIP bias, which now have cooled, and exhibit photospheric abundances everywhere, except at their looptops.

In conclusion, we were able to infer the spatially resolved SPICE measurements of rapid FIP-bias evolution in post-flare loop plasma. The results show the complexity of the phenomenon. The comparison between SO/SPICE and XSM shows the clear discrepancy of the plasma composition evolution. In particular, we suggest that the contrasting behavior arises from differing regions emitting the different radiation detected by the two instruments. We suggest this by using the lowest energy channel of the STIX instrument, which observed the event from the SO vantage. This shows the potential of SPICE for observing composition changes during AR flares and provides spatiotemporal constraints of the FIP fractionation process. 

\begin{acknowledgments}
We would like to thank the referee for the suggestions which significantly improved the manuscript.
Solar Orbiter is a space mission of international collaboration between ESA and NASA, operated by ESA. The development of the
SPICE instrument was funded by ESA and ESA member states (France, Germany, Norway, Switzerland, United Kingdom). The SPICE hardware consortium was led by Science and Technology Facilities Council (STFC) RAL Space
and included Institut d’Astrophysique Spatiale (IAS), Max-Planck-Institut f\"{u}r
Sonnensystemforschung (MPS), Physikalisch-Meteorologisches Observatorium
Davos and World Radiation Center (PMOD/WRC), Institute of Theoretical Astrophysics (University of Oslo), NASA Goddard Space Flight Center (GSFC)
and Southwest Research Institute (SwRI). The efforts at SwRI for Solar Orbiter
SPICE are supported by NASA under GSFC subcontract 80GSFC20C0053 to
Southwest Research Institute.
We acknowledge the use of data from the Solar X-ray Monitor (XSM) on board the Chandrayaan-2 mission of the Indian Space Research Organisation (ISRO), archived at the Indian Space Science Data Centre (ISSDC). XSM was developed by Physical Research Laboratory (PRL) with support from various ISRO centers. 
CHIANTI is a collaborative project involving George Mason University, the University of Michigan (USA), University of Cambridge (UK) and NASA Goddard Space Flight Center (USA). 
R.P. is supported by NASA Solar Orbiter Guest Investigator Grant number 80NSSC24K1243. R.J.F. thanks support from NASA HGI award 80NSSC25K7927. L.P.C. gratefully acknowledges funding by the European Union (grant agreement No 101039844); views and opinions expressed are however those of the author(s) only and do not necessarily reflect those of the European Union or the European Research Council. A.C. was supported by NASA HFORT award 80NSSC22M0111.
\end{acknowledgments}

%
\facilities{Solar Orbiter~\citep{Muller2020}: SPICE~\citep{2020A&A...642A..14S} and EUI~\citep{Rochus2020}; SDO/HMI~\citep{HMI_2012},
SDO/AIA~\citep{2012SoPh..275...17L}, Hinode/EIS~\citep{2007SoPh..243...19C}}, Chandray\textbf{a}an-2 XSM~\citep{VADAWALE20142021}, GOES-R/EXIS~\citep{goesr_exis_l2}.

\software{numpy~\citep{numpy2020}, 
          scipy~\citep{Scipy2020}, 
          astropy~\citep{2022ApJ...935..167A}, 
          sunpy~\citep{sunpy_community2020}, 
          fiasco~\citep{Barnes2024}, 
          CHIANTI~\citep{Dere97,ChiantiPy},
 EMToolkit~\citep{Plowman_EMToolKit_A_Standardized}, \emph{cmcrameri} colormaps~\citep{zenodo_8409685}.} 

\bibliography{sample7}{}

@ARTICLE{2014ApJ...786L...2W,
	author = {{Warren}, Harry P.},
	title = "{Measurements of Absolute Abundances in Solar Flares}",
	journal = {\apjl},
	year = 2014,
	month = may,
	volume = {786},
	number = {1},
	eid = {L2},
	pages = {L2},
	doi = {10.1088/2041-8205/786/1/L2},
	archivePrefix = {arXiv},
	eprint = {1310.4765},
	primaryClass = {astro-ph.SR},
	adsurl = {https://ui.adsabs.harvard.edu/abs/2014ApJ...786L...2W}
}

@ARTICLE{2022ApJ...935..167A,
	author = {{Astropy Collaboration} and {Price-Whelan}, Adrian M. and {Lim}, Pey Lian and {Earl}, Nicholas and {Starkman}, Nathaniel and {Bradley}, Larry and {Shupe}, David L. and {Patil}, Aarya A. and {Corrales}, Lia and {Brasseur}, C.~E. and {N{\"o}the}, Maximilian and {Donath}, Axel and {Tollerud}, Erik and {Morris}, Brett M. and {Ginsburg}, Adam and {Vaher}, Eero and {Weaver}, Benjamin A. and {Tocknell}, James and {Jamieson}, William and {van Kerkwijk}, Marten H. and {Robitaille}, Thomas P. and {Merry}, Bruce and {Bachetti}, Matteo and {G{\"u}nther}, H. Moritz and {Aldcroft}, Thomas L. and {Alvarado-Montes}, Jaime A. and {Archibald}, Anne M. and {B{\'o}di}, Attila and {Bapat}, Shreyas and {Barentsen}, Geert and {Baz{\'a}n}, Juanjo and {Biswas}, Manish and {Boquien}, M{\'e}d{\'e}ric and {Burke}, D.~J. and {Cara}, Daria and {Cara}, Mihai and {Conroy}, Kyle E. and {Conseil}, Simon and {Craig}, Matthew W. and {Cross}, Robert M. and {Cruz}, Kelle L. and {D'Eugenio}, Francesco and {Dencheva}, Nadia and {Devillepoix}, Hadrien A.~R. and {Dietrich}, J{\"o}rg P. and {Eigenbrot}, Arthur Davis and {Erben}, Thomas and {Ferreira}, Leonardo and {Foreman-Mackey}, Daniel and {Fox}, Ryan and {Freij}, Nabil and {Garg}, Suyog and {Geda}, Robel and {Glattly}, Lauren and {Gondhalekar}, Yash and {Gordon}, Karl D. and {Grant}, David and {Greenfield}, Perry and {Groener}, Austen M. and {Guest}, Steve and {Gurovich}, Sebastian and {Handberg}, Rasmus and {Hart}, Akeem and {Hatfield-Dodds}, Zac and {Homeier}, Derek and {Hosseinzadeh}, Griffin and {Jenness}, Tim and {Jones}, Craig K. and {Joseph}, Prajwel and {Kalmbach}, J. Bryce and {Karamehmetoglu}, Emir and {Ka{\l}uszy{\'n}ski}, Miko{\l}aj and {Kelley}, Michael S.~P. and {Kern}, Nicholas and {Kerzendorf}, Wolfgang E. and {Koch}, Eric W. and {Kulumani}, Shankar and {Lee}, Antony and {Ly}, Chun and {Ma}, Zhiyuan and {MacBride}, Conor and {Maljaars}, Jakob M. and {Muna}, Demitri and {Murphy}, N.~A. and {Norman}, Henrik and {O'Steen}, Richard and {Oman}, Kyle A. and {Pacifici}, Camilla and {Pascual}, Sergio and {Pascual-Granado}, J. and {Patil}, Rohit R. and {Perren}, Gabriel I. and {Pickering}, Timothy E. and {Rastogi}, Tanuj and {Roulston}, Benjamin R. and {Ryan}, Daniel F. and {Rykoff}, Eli S. and {Sabater}, Jose and {Sakurikar}, Parikshit and {Salgado}, Jes{\'u}s and {Sanghi}, Aniket and {Saunders}, Nicholas and {Savchenko}, Volodymyr and {Schwardt}, Ludwig and {Seifert-Eckert}, Michael and {Shih}, Albert Y. and {Jain}, Anany Shrey and {Shukla}, Gyanendra and {Sick}, Jonathan and {Simpson}, Chris and {Singanamalla}, Sudheesh and {Singer}, Leo P. and {Singhal}, Jaladh and {Sinha}, Manodeep and {Sip{\H{o}}cz}, Brigitta M. and {Spitler}, Lee R. and {Stansby}, David and {Streicher}, Ole and {{\v{S}}umak}, Jani and {Swinbank}, John D. and {Taranu}, Dan S. and {Tewary}, Nikita and {Tremblay}, Grant R. and {de Val-Borro}, Miguel and {Van Kooten}, Samuel J. and {Vasovi{\'c}}, Zlatan and {Verma}, Shresth and {de Miranda Cardoso}, Jos{\'e} Vin{\'\i}cius and {Williams}, Peter K.~G. and {Wilson}, Tom J. and {Winkel}, Benjamin and {Wood-Vasey}, W.~M. and {Xue}, Rui and {Yoachim}, Peter and {Zhang}, Chen and {Zonca}, Andrea and {Astropy Project Contributors}},
	title = "{The Astropy Project: Sustaining and Growing a Community-oriented Open-source Project and the Latest Major Release (v5.0) of the Core Package}",
	journal = {\apj},
	year = 2022,
	month = aug,
	volume = {935},
	number = {2},
	eid = {167},
	pages = {167},
	doi = {10.3847/1538-4357/ac7c74},
	archivePrefix = {arXiv},
	eprint = {2206.14220},
	primaryClass = {astro-ph.IM},
	adsurl = {https://ui.adsabs.harvard.edu/abs/2022ApJ...935..167A}
}

@ARTICLE{Varesano_2025,
	author = {{Varesano}, T. and {Hassler}, D.~M. and {Zambrana Prado}, N. and {Laming}, J.~M. and {Plowman}, J. and {Knipp}, D.~J. and {Molnar}, M. and {Barczynski}, K. and {The Spice Consortium}},
	title = "{First ionization potential bias evolution in an emerging active region as observed in SPICE synoptic observations}",
	journal = {\aap},
	year = 2026,
	month = feb,
	volume = {706},
	eid = {A155},
	pages = {A155},
	doi = {10.1051/0004-6361/202554166},
	archivePrefix = {arXiv},
	eprint = {2502.12045},
	primaryClass = {astro-ph.SR},
	adsurl = {https://ui.adsabs.harvard.edu/abs/2026A&A...706A.155V}
}

@software{Barnes2024,
	author       = {Will Barnes and
	David Stansby and
	Nick Murphy and
	Jeffrey Reep and
	Laura Hayes and
	Stuart Mumford and
	Nabil Freij},
	title        = {{fiasco}},
	month        = jan,
	year         = 2025,
	publisher    = {Zenodo},
	version      = {v0.4.0},
	doi          = {10.5281/zenodo.14757042},
	url          = {https://doi.org/10.5281/zenodo.14757042}
}

@ARTICLE{2026A&A...705A.113C,
	author = {{Chitta}, L.~P. and {Pontin}, D.~I. and {Priest}, E.~R. and {Berghmans}, D. and {Kraaikamp}, E. and {Rodriguez}, L. and {Verbeeck}, C. and {Zhukov}, A.~N. and {Krucker}, S. and {Aznar Cuadrado}, R. and {Calchetti}, D. and {Hirzberger}, J. and {Peter}, H. and {Sch{\"u}hle}, U. and {Solanki}, S.~K. and {Teriaca}, L. and {Giunta}, A.~S. and {Auch{\`e}re}, F. and {Harra}, L. and {M{\"u}ller}, D.},
	title = "{A magnetic avalanche as the central engine powering a solar flare}",
	journal = {\aap},
	year = 2026,
	month = jan,
	volume = {705},
	eid = {A113},
	pages = {A113},
	doi = {10.1051/0004-6361/202557253},
	archivePrefix = {arXiv},
	eprint = {2503.12235},
	primaryClass = {astro-ph.SR},
	adsurl = {https://ui.adsabs.harvard.edu/abs/2026A&A...705A.113C}
}

@ARTICLE{sunpy_community2020,
	doi = {10.3847/1538-4357/ab4f7a},
	url = {https://iopscience.iop.org/article/10.3847/1538-4357/ab4f7a},
	author = {{The SunPy Community} and Barnes, Will T. and Bobra, Monica G. and Christe, Steven D. and Freij, Nabil and Hayes, Laura A. and Ireland, Jack and Mumford, Stuart and Perez-Suarez, David and Ryan, Daniel F. and Shih, Albert Y. and Chanda, Prateek and Glogowski, Kolja and Hewett, Russell and Hughitt, V. Keith and Hill, Andrew and Hiware, Kaustubh and Inglis, Andrew and Kirk, Michael S. F. and Konge, Sudarshan and Mason, James Paul and Maloney, Shane Anthony and Murray, Sophie A. and Panda, Asish and Park, Jongyeob and Pereira, Tiago M. D. and Reardon, Kevin and Savage, Sabrina and Sipőcz, Brigitta M. and Stansby, David and Jain, Yash and Taylor, Garrison and Yadav, Tannmay and Rajul and Dang, Trung Kien},
	title = {The SunPy Project: Open Source Development and Status of the Version 1.0 Core Package},
	journal = {The Astrophysical Journal},
	volume = {890},
	issue = {1},
	pages = {68-},
	publisher = {American Astronomical Society},
	year = {2020}
}

@ARTICLE{Scipy2020,
	author  = {Virtanen, Pauli and Gommers, Ralf and Oliphant, Travis E. and
	Haberland, Matt and Reddy, Tyler and Cournapeau, David and
	Burovski, Evgeni and Peterson, Pearu and Weckesser, Warren and
	Bright, Jonathan and {van der Walt}, St{\'e}fan J. and
	Brett, Matthew and Wilson, Joshua and Millman, K. Jarrod and
	Mayorov, Nikolay and Nelson, Andrew R. J. and Jones, Eric and
	Kern, Robert and Larson, Eric and Carey, C J and
	Polat, {\.I}lhan and Feng, Yu and Moore, Eric W. and
	{VanderPlas}, Jake and Laxalde, Denis and Perktold, Josef and
	Cimrman, Robert and Henriksen, Ian and Quintero, E. A. and
	Harris, Charles R. and Archibald, Anne M. and
	Ribeiro, Ant{\^o}nio H. and Pedregosa, Fabian and
	{van Mulbregt}, Paul and {SciPy 1.0 Contributors}},
	title   = {{{SciPy} 1.0: Fundamental Algorithms for Scientific
	Computing in Python}},
	journal = {Nature Methods},
	year    = {2020},
	volume  = {17},
	pages   = {261--272},
	adsurl  = {https://rdcu.be/b08Wh},
	doi     = {10.1038/s41592-019-0686-2},
}

@ARTICLE{2021A&A...653A.141A,
	author = {{Asplund}, M. and {Amarsi}, A.~M. and {Grevesse}, N.},
	title = "{The chemical make-up of the Sun: A 2020 vision}",
	journal = {\aap},
	year = 2021,
	month = sep,
	volume = {653},
	eid = {A141},
	pages = {A141},
	doi = {10.1051/0004-6361/202140445},
	archivePrefix = {arXiv},
	eprint = {2105.01661},
	primaryClass = {astro-ph.SR},
	adsurl = {https://ui.adsabs.harvard.edu/abs/2021A&A...653A.141A}
}

@ARTICLE{2020A&A...642A..14S,
	author = {{SPICE Consortium} and {Anderson}, M. and {Appourchaux}, T. and {Auch{\`e}re}, F. and {Aznar Cuadrado}, R. and {Barbay}, J. and {Baudin}, F. and {Beardsley}, S. and {Bocchialini}, K. and {Borgo}, B. and {Bruzzi}, D. and {Buchlin}, E. and {Burton}, G. and {B{\"u}chel}, V. and {Caldwell}, M. and {Caminade}, S. and {Carlsson}, M. and {Curdt}, W. and {Davenne}, J. and {Davila}, J. and {Deforest}, C.~E. and {Del Zanna}, G. and {Drummond}, D. and {Dubau}, J. and {Dumesnil}, C. and {Dunn}, G. and {Eccleston}, P. and {Fludra}, A. and {Fredvik}, T. and {Gabriel}, A. and {Giunta}, A. and {Gottwald}, A. and {Griffin}, D. and {Grundy}, T. and {Guest}, S. and {Gyo}, M. and {Haberreiter}, M. and {Hansteen}, V. and {Harrison}, R. and {Hassler}, D.~M. and {Haugan}, S.~V.~H. and {Howe}, C. and {Janvier}, M. and {Klein}, R. and {Koller}, S. and {Kucera}, T.~A. and {Kouliche}, D. and {Marsch}, E. and {Marshall}, A. and {Marshall}, G. and {Matthews}, S.~A. and {McQuirk}, C. and {Meining}, S. and {Mercier}, C. and {Morris}, N. and {Morse}, T. and {Munro}, G. and {Parenti}, S. and {Pastor-Santos}, C. and {Peter}, H. and {Pfiffner}, D. and {Phelan}, P. and {Philippon}, A. and {Richards}, A. and {Rogers}, K. and {Sawyer}, C. and {Schlatter}, P. and {Schmutz}, W. and {Sch{\"u}hle}, U. and {Shaughnessy}, B. and {Sidher}, S. and {Solanki}, S.~K. and {Speight}, R. and {Spescha}, M. and {Szwec}, N. and {Tamiatto}, C. and {Teriaca}, L. and {Thompson}, W. and {Tosh}, I. and {Tustain}, S. and {Vial}, J. -C. and {Walls}, B. and {Waltham}, N. and {Wimmer-Schweingruber}, R. and {Woodward}, S. and {Young}, P. and {de Groof}, A. and {Pacros}, A. and {Williams}, D. and {M{\"u}ller}, D.},
	title = "{The Solar Orbiter SPICE instrument. An extreme UV imaging spectrometer}",
	journal = {\aap},
	year = 2020,
	month = oct,
	volume = {642},
	eid = {A14},
	pages = {A14},
	doi = {10.1051/0004-6361/201935574},
	archivePrefix = {arXiv},
	eprint = {1909.01183},
	primaryClass = {astro-ph.IM},
	adsurl = {https://ui.adsabs.harvard.edu/abs/2020A&A...642A..14S}
}

@ARTICLE{Plowman2020,
	author = {{Plowman}, Joseph and {Caspi}, Amir},
	title = "{A Fast, Simple, Robust Algorithm for Coronal Temperature Reconstruction}",
	journal = {\apj},
	year = 2020,
	month = dec,
	volume = {905},
	number = {1},
	eid = {17},
	pages = {17},
	doi = {10.3847/1538-4357/abc260},
	archivePrefix = {arXiv},
	eprint = {2006.06828},
	primaryClass = {astro-ph.SR},
	adsurl = {https://ui.adsabs.harvard.edu/abs/2020ApJ...905...17P}
}

@ARTICLE{Rochus2020,
	author = {{Rochus}, P. and {Auch{\`e}re}, F. and {Berghmans}, D. and {Harra}, L. and {Schmutz}, W. and {Sch{\"u}hle}, U. and {Addison}, P. and {Appourchaux}, T. and {Aznar Cuadrado}, R. and {Baker}, D. and {Barbay}, J. and {Bates}, D. and {BenMoussa}, A. and {Bergmann}, M. and {Beurthe}, C. and {Borgo}, B. and {Bonte}, K. and {Bouzit}, M. and {Bradley}, L. and {B{\"u}chel}, V. and {Buchlin}, E. and {B{\"u}chner}, J. and {Cab{\'e}}, F. and {Cadiergues}, L. and {Chaigneau}, M. and {Chares}, B. and {Choque Cortez}, C. and {Coker}, P. and {Condamin}, M. and {Coumar}, S. and {Curdt}, W. and {Cutler}, J. and {Davies}, D. and {Davison}, G. and {Defise}, J. -M. and {Del Zanna}, G. and {Delmotte}, F. and {Delouille}, V. and {Dolla}, L. and {Dumesnil}, C. and {D{\"u}rig}, F. and {Enge}, R. and {Fran{\c{c}}ois}, S. and {Fourmond}, J. -J. and {Gillis}, J. -M. and {Giordanengo}, B. and {Gissot}, S. and {Green}, L.~M. and {Guerreiro}, N. and {Guilbaud}, A. and {Gyo}, M. and {Haberreiter}, M. and {Hafiz}, A. and {Hailey}, M. and {Halain}, J. -P. and {Hansotte}, J. and {Hecquet}, C. and {Heerlein}, K. and {Hellin}, M. -L. and {Hemsley}, S. and {Hermans}, A. and {Hervier}, V. and {Hochedez}, J. -F. and {Houbrechts}, Y. and {Ihsan}, K. and {Jacques}, L. and {J{\'e}r{\^o}me}, A. and {Jones}, J. and {Kahle}, M. and {Kennedy}, T. and {Klaproth}, M. and {Kolleck}, M. and {Koller}, S. and {Kotsialos}, E. and {Kraaikamp}, E. and {Langer}, P. and {Lawrenson}, A. and {Le Clech'}, J. -C. and {Lenaerts}, C. and {Liebecq}, S. and {Linder}, D. and {Long}, D.~M. and {Mampaey}, B. and {Markiewicz-Innes}, D. and {Marquet}, B. and {Marsch}, E. and {Matthews}, S. and {Mazy}, E. and {Mazzoli}, A. and {Meining}, S. and {Meltchakov}, E. and {Mercier}, R. and {Meyer}, S. and {Monecke}, M. and {Monfort}, F. and {Morinaud}, G. and {Moron}, F. and {Mountney}, L. and {M{\"u}ller}, R. and {Nicula}, B. and {Parenti}, S. and {Peter}, H. and {Pfiffner}, D. and {Philippon}, A. and {Phillips}, I. and {Plesseria}, J. -Y. and {Pylyser}, E. and {Rabecki}, F. and {Ravet-Krill}, M. -F. and {Rebellato}, J. and {Renotte}, E. and {Rodriguez}, L. and {Roose}, S. and {Rosin}, J. and {Rossi}, L. and {Roth}, P. and {Rouesnel}, F. and {Roulliay}, M. and {Rousseau}, A. and {Ruane}, K. and {Scanlan}, J. and {Schlatter}, P. and {Seaton}, D.~B. and {Silliman}, K. and {Smit}, S. and {Smith}, P.~J. and {Solanki}, S.~K. and {Spescha}, M. and {Spencer}, A. and {Stegen}, K. and {Stockman}, Y. and {Szwec}, N. and {Tamiatto}, C. and {Tandy}, J. and {Teriaca}, L. and {Theobald}, C. and {Tychon}, I. and {van Driel-Gesztelyi}, L. and {Verbeeck}, C. and {Vial}, J. -C. and {Werner}, S. and {West}, M.~J. and {Westwood}, D. and {Wiegelmann}, T. and {Willis}, G. and {Winter}, B. and {Zerr}, A. and {Zhang}, X. and {Zhukov}, A.~N.},
	title = "{The Solar Orbiter EUI instrument: The Extreme Ultraviolet Imager}",
	journal = {\aap},
	year = 2020,
	month = oct,
	volume = {642},
	eid = {A8},
	pages = {A8},
	doi = {10.1051/0004-6361/201936663},
	adsurl = {https://ui.adsabs.harvard.edu/abs/2020A&A...642A...8R}
}

@ARTICLE{Muller2020,
	author = {{M{\"u}ller}, D. and {St. Cyr}, O.~C. and {Zouganelis}, I. and {Gilbert}, H.~R. and {Marsden}, R. and {Nieves-Chinchilla}, T. and {Antonucci}, E. and {Auch{\`e}re}, F. and {Berghmans}, D. and {Horbury}, T.~S. and {Howard}, R.~A. and {Krucker}, S. and {Maksimovic}, M. and {Owen}, C.~J. and {Rochus}, P. and {Rodriguez-Pacheco}, J. and {Romoli}, M. and {Solanki}, S.~K. and {Bruno}, R. and {Carlsson}, M. and {Fludra}, A. and {Harra}, L. and {Hassler}, D.~M. and {Livi}, S. and {Louarn}, P. and {Peter}, H. and {Sch{\"u}hle}, U. and {Teriaca}, L. and {del Toro Iniesta}, J.~C. and {Wimmer-Schweingruber}, R.~F. and {Marsch}, E. and {Velli}, M. and {De Groof}, A. and {Walsh}, A. and {Williams}, D.},
	title = "{The Solar Orbiter mission. Science overview}",
	journal = {\aap},
	year = 2020,
	month = oct,
	volume = {642},
	eid = {A1},
	pages = {A1},
	doi = {10.1051/0004-6361/202038467},
	archivePrefix = {arXiv},
	eprint = {2009.00861},
	primaryClass = {astro-ph.SR},
	adsurl = {https://ui.adsabs.harvard.edu/abs/2020A&A...642A...1M}
}

@Article{numpy2020,
	title         = {Array programming with {NumPy}},
	author        = {Charles R. Harris and K. Jarrod Millman and St{\'{e}}fan J.
	van der Walt and Ralf Gommers and Pauli Virtanen and David
	Cournapeau and Eric Wieser and Julian Taylor and Sebastian
	Berg and Nathaniel J. Smith and Robert Kern and Matti Picus
	and Stephan Hoyer and Marten H. van Kerkwijk and Matthew
	Brett and Allan Haldane and Jaime Fern{\'{a}}ndez del
	R{\'{i}}o and Mark Wiebe and Pearu Peterson and Pierre
	G{\'{e}}rard-Marchant and Kevin Sheppard and Tyler Reddy and
	Warren Weckesser and Hameer Abbasi and Christoph Gohlke and
	Travis E. Oliphant},
	year          = {2020},
	month         = sep,
	journal       = {Nature},
	volume        = {585},
	number        = {7825},
	pages         = {357--362},
	doi           = {10.1038/s41586-020-2649-2},
	publisher     = {Springer Science and Business Media {LLC}},
	url           = {https://doi.org/10.1038/s41586-020-2649-2}
}

@ARTICLE{Plowman_2023,
	author = {{Plowman}, J.~E. and {Hassler}, D.~M. and {Auch{\`e}re}, F. and {Aznar Cuadrado}, R. and {Fludra}, A. and {Mandal}, S. and {Peter}, H.},
	title = "{SPICE point spread function correction: General framework and capability demonstration}",
	journal = {\aap},
	year = 2023,
	month = oct,
	volume = {678},
	eid = {A52},
	pages = {A52},
	doi = {10.1051/0004-6361/202245582},
	archivePrefix = {arXiv},
	eprint = {2211.16635},
	primaryClass = {astro-ph.SR},
	adsurl = {https://ui.adsabs.harvard.edu/abs/2023A&A...678A..52P}
}

@ARTICLE{Gieseler_SolarMACH_2022,
	author = {{Gieseler}, Jan and {Dresing}, Nina and {Palmroos}, Christian and {Freiherr von Forstner}, Johan L. and {Price}, Daniel J. and {Vainio}, Rami and {Kouloumvakos}, Athanasios and {Rodr{\'\i}guez-Garc{\'\i}a}, Laura and {Trotta}, Domenico and {G{\'e}not}, Vincent and {Masson}, Arnaud and {Roth}, Markus and {Veronig}, Astrid},
	title = "{Solar-MACH: An open-source tool to analyze solar magnetic connection configurations}",
	journal = {Frontiers in Astronomy and Space Sciences},
	year = 2023,
	month = feb,
	volume = {9},
	eid = {384},
	pages = {384},
	doi = {10.3389/fspas.2022.1058810},
	archivePrefix = {arXiv},
	eprint = {2210.00819},
	primaryClass = {astro-ph.SR},
	adsurl = {https://ui.adsabs.harvard.edu/abs/2023FrASS...958810G}
}

@ARTICLE{2026arXiv260504223Y,
	author = {{Young}, Peter R. and {Mondal}, Biswajit},
	title = "{Modeling Flare Continuum Emission Observed by Hinode/EIS: Instrument Calibration and Element Composition Results}",
	journal = {arXiv e-prints},
	year = 2026,
	month = may,
	eid = {arXiv:2605.04223},
	pages = {arXiv:2605.04223},
	archivePrefix = {arXiv},
	eprint = {2605.04223},
	primaryClass = {astro-ph.SR},
	adsurl = {https://ui.adsabs.harvard.edu/abs/2026arXiv260504223Y}
}

@software{Plowman_EMToolKit_A_Standardized,
	author = {Plowman, Joseph and Barnes, Will T. and Caspi, Amir and Cheung, Mark and Gilly, C.}, 
	year = 2025,
	title = {{EMToolKit: A Standardized Framework for Computing and Visualizing Differential Emission Measures}}
}

@software{ChiantiPy,
	author = {{Dere}, Ken},
	title = "{ChiantiPy: Python package for the CHIANTI atomic database}",
	howpublished = {Astrophysics Source Code Library, record ascl:1308.017},
	year = 2013,
	month = aug,
	eid = {ascl:1308.017},
	adsurl = {https://ui.adsabs.harvard.edu/abs/2013ascl.soft08017D}
}

@INPROCEEDINGS{1996ASPC..101...17A,
	author = {{Arnaud}, K.~A.},
	title = "{XSPEC: The First Ten Years}",
	booktitle = {Astronomical Data Analysis Software and Systems V},
	year = 1996,
	editor = {{Jacoby}, George H. and {Barnes}, Jeannette},
	series = {Astronomical Society of the Pacific Conference Series},
	volume = {101},
	month = jan,
	pages = {17},
	adsurl = {https://ui.adsabs.harvard.edu/abs/1996ASPC..101...17A}
}

@ARTICLE{2021ApJ...920....4M,
	author = {{Mondal}, Biswajit and {Sarkar}, Aveek and {Vadawale}, Santosh V. and {Mithun}, N.~P.~S. and {Janardhan}, P. and {Del Zanna}, Giulio and {Mason}, Helen E. and {Mitra-Kraev}, Urmila and {Narendranath}, S.},
	title = "{Evolution of Elemental Abundances during B-Class Solar Flares: Soft X-Ray Spectral Measurements with Chandrayaan-2 XSM}",
	journal = {\apj},
	year = 2021,
	month = oct,
	volume = {920},
	number = {1},
	eid = {4},
	pages = {4},
	doi = {10.3847/1538-4357/ac14c1},
	archivePrefix = {arXiv},
	eprint = {2107.07825},
	primaryClass = {astro-ph.SR},
	adsurl = {https://ui.adsabs.harvard.edu/abs/2021ApJ...920....4M}
}

@ARTICLE{2024ApJ...976..188B,
	author = {{Brooks}, David H. and {Warren}, Harry P. and {Baker}, Deborah and {Matthews}, Sarah A. and {Yardley}, Stephanie L.},
	title = "{An Elemental Abundance Diagnostic for Coordinated Solar Orbiter/SPICE and Hinode/EIS Observations}",
	journal = {\apj},
	year = 2024,
	month = dec,
	volume = {976},
	number = {2},
	eid = {188},
	pages = {188},
	doi = {10.3847/1538-4357/ad87ef},
	archivePrefix = {arXiv},
	eprint = {2410.15606},
	primaryClass = {astro-ph.SR},
	adsurl = {https://ui.adsabs.harvard.edu/abs/2024ApJ...976..188B}
}

@ARTICLE{1991AdSpR..11a.269M,
	author = {{Meyer}, Jean-Paul},
	title = "{Diagnostic methods for coronal abundances}",
	journal = {Advances in Space Research},
	year = 1991,
	month = jan,
	volume = {11},
	number = {1},
	pages = {269-280},
	doi = {10.1016/0273-1177(91)90120-9},
	adsurl = {https://ui.adsabs.harvard.edu/abs/1991AdSpR..11a.269M}
}

@article{Doschek_2015,
	doi = {10.1088/2041-8205/808/1/L7},
	url = {https://doi.org/10.1088/2041-8205/808/1/L7},
	year = {2015},
	month = {jul},
	publisher = {The American Astronomical Society},
	volume = {808},
	number = {1},
	pages = {L7},
	author = {Doschek, G. A. and Warren, H. P. and Feldman, U.},
	title = {ANOMALOUS RELATIVE AR/CA CORONAL ABUNDANCES OBSERVED BY THE HINODE/EUV IMAGING SPECTROMETER NEAR SUNSPOTS},
	journal = {The Astrophysical Journal Letters}
}

@ARTICLE{2009ApJ...695..954L,
	author = {{Laming}, J. Martin},
	title = "{Non-Wkb Models of the First Ionization Potential Effect: Implications for Solar Coronal Heating and the Coronal Helium and Neon Abundances}",
	journal = {\apj},
	year = 2009,
	month = apr,
	volume = {695},
	number = {2},
	pages = {954-969},
	doi = {10.1088/0004-637X/695/2/954},
	archivePrefix = {arXiv},
	eprint = {0901.3350},
	primaryClass = {astro-ph.SR},
	adsurl = {https://ui.adsabs.harvard.edu/abs/2009ApJ...695..954L}
}

@article{Martínez-Sykora_2023,
	doi = {10.3847/1538-4357/acc465},
	url = {https://doi.org/10.3847/1538-4357/acc465},
	year = {2023},
	month = {jun},
	publisher = {The American Astronomical Society},
	volume = {949},
	number = {2},
	pages = {112},
	author = {Martínez-Sykora, Juan and De Pontieu, Bart and Hansteen, Viggo H. and Testa, Paola and Wargnier, Q. M. and Szydlarski, Mikolaj},
	title = {The Impact of Multifluid Effects in the Solar Chromosphere on the Ponderomotive Force under SE and NEQ Ionization Conditions},
	journal = {The Astrophysical Journal}
}

@ARTICLE{2021FrASS...8....2R,
	author = {{R{\'e}ville}, Victor and {Rouillard}, Alexis P. and {Velli}, Marco and {Verdini}, Andrea and {Buchlin}, {\'E}ric and {Lavarra}, Michael and {Poirier}, Nicolas},
	title = "{Investigating the origin of the FIP effect with a shell turbulence model}",
	journal = {Frontiers in Astronomy and Space Sciences},
	year = 2021,
	month = feb,
	volume = {8},
	eid = {2},
	pages = {2},
	doi = {10.3389/fspas.2021.619463},
	archivePrefix = {arXiv},
	eprint = {2101.01440},
	primaryClass = {astro-ph.SR},
	adsurl = {https://ui.adsabs.harvard.edu/abs/2021FrASS...8....2R}
}

@article{Ko_2016,
	doi = {10.3847/0004-637X/826/2/126},
	url = {https://doi.org/10.3847/0004-637X/826/2/126},
	year = {2016},
	month = {jul},
	publisher = {The American Astronomical Society},
	volume = {826},
	number = {2},
	pages = {126},
	author = {Ko, Yuan-Kuen and Young, Peter R. and Muglach, Karin and Warren, Harry P. and Ugarte-Urra, Ignacio},
	title = {CORRELATION OF CORONAL PLASMA PROPERTIES AND SOLAR MAGNETIC FIELD IN A DECAYING ACTIVE REGION},
	journal = {The Astrophysical Journal}
}

@ARTICLE{2023ApJ...959...72M,
	author = {{Mihailescu}, Teodora and {Brooks}, David H. and {Laming}, J. Martin and {Baker}, Deborah and {Green}, Lucie M. and {James}, Alexander W. and {Long}, David M. and {van Driel-Gesztelyi}, Lidia and {Stangalini}, Marco},
	title = "{Intriguing Plasma Composition Pattern in a Solar Active Region: A Result of Nonresonant Alfv{\'e}n Waves?}",
	journal = {\apj},
	year = 2023,
	month = dec,
	volume = {959},
	number = {2},
	eid = {72},
	pages = {72},
	doi = {10.3847/1538-4357/ad05bf},
	archivePrefix = {arXiv},
	eprint = {2310.13677},
	primaryClass = {astro-ph.SR},
	adsurl = {https://ui.adsabs.harvard.edu/abs/2023ApJ...959...72M}
}

@ARTICLE{2021ApJ...909...17L,
	author = {{Laming}, J. Martin},
	title = "{The FIP and Inverse-FIP Effects in Solar Flares}",
	journal = {\apj},
	year = 2021,
	month = mar,
	volume = {909},
	number = {1},
	eid = {17},
	pages = {17},
	doi = {10.3847/1538-4357/abd9c3},
	archivePrefix = {arXiv},
	eprint = {2101.03038},
	primaryClass = {astro-ph.SR},
	adsurl = {https://ui.adsabs.harvard.edu/abs/2021ApJ...909...17L}
}

@ARTICLE{2011ApJ...727L..13B,
	author = {{Brooks}, David H. and {Warren}, Harry P.},
	title = "{Establishing a Connection Between Active Region Outflows and the Solar Wind: Abundance Measurements with EIS/Hinode}",
	journal = {\apjl},
	year = 2011,
	month = jan,
	volume = {727},
	number = {1},
	eid = {L13},
	pages = {L13},
	doi = {10.1088/2041-8205/727/1/L13},
	archivePrefix = {arXiv},
	eprint = {1009.4291},
	primaryClass = {astro-ph.SR},
	adsurl = {https://ui.adsabs.harvard.edu/abs/2011ApJ...727L..13B}
}

@misc{goesr_exis_l2,
	author       = {{Machol}, Janet and {Codrescu}, Stefan and {Viereck}, Rodney},
	title        = {{GOES-R Series Extreme Ultraviolet and X-ray Irradiance Sensors (EXIS) Level 2 Products}},
	year         = {2018},
	publisher    = {{NOAA National Centers for Environmental Information}},
	doi          = {10.25921/94P8-YE57},
}

@ARTICLE{2020SoPh..295..175N,
	author = {{Narendranath}, Shyama and {Sreekumar}, P. and {Pillai}, Netra S. and {Panini}, Singam and {Sankarasubramanian}, K. and {Huovelin}, Juhani},
	title = "{Coronal Elemental Abundance: New Results from Soft X-Ray Spectroscopy of the Sun}",
	journal = {\solphys},
	year = 2020,
	month = dec,
	volume = {295},
	number = {12},
	eid = {175},
	pages = {175},
	doi = {10.1007/s11207-020-01738-5},
	archivePrefix = {arXiv},
	eprint = {2011.08584},
	primaryClass = {astro-ph.SR},
	adsurl = {https://ui.adsabs.harvard.edu/abs/2020SoPh..295..175N}
}

@ARTICLE{2019ApJ...875...35B,
	author = {{Baker}, Deborah and {van Driel-Gesztelyi}, Lidia and {Brooks}, David H. and {Valori}, Gherardo and {James}, Alexander W. and {Laming}, J. Martin and {Long}, David M. and {D{\'e}moulin}, Pascal and {Green}, Lucie M. and {Matthews}, Sarah A. and {Ol{\'a}h}, Katalin and {K{\H{o}}v{\'a}ri}, Zsolt},
	title = "{Transient Inverse-FIP Plasma Composition Evolution within a Solar Flare}",
	journal = {\apj},
	year = 2019,
	month = apr,
	volume = {875},
	number = {1},
	eid = {35},
	pages = {35},
	doi = {10.3847/1538-4357/ab07c1},
	archivePrefix = {arXiv},
	eprint = {1902.06948},
	primaryClass = {astro-ph.SR},
	adsurl = {https://ui.adsabs.harvard.edu/abs/2019ApJ...875...35B}
}

@ARTICLE{1985ApJS...57..173M,
	author = {{Meyer}, J.-P.},
	title = "{Solar-stellar outer atmospheres and energetic particles, and galactic cosmic rays}",
	journal = {\apjs},
	year = 1985,
	month = jan,
	volume = {57},
	pages = {173-204},
	doi = {10.1086/191001},
	adsurl = {https://ui.adsabs.harvard.edu/abs/1985ApJS...57..173M}
}

@ARTICLE{1992ApJS...81..387F,
	author = {{Feldman}, U. and {Mandelbaum}, P. and {Seely}, J.~F. and {Doschek}, G.~A. and {Gursky}, H.},
	title = "{The Potential for Plasma Diagnostics from Stellar Extreme-Ultraviolet Observations}",
	journal = {\apjs},
	year = 1992,
	month = jul,
	volume = {81},
	pages = {387},
	doi = {10.1086/191698},
	adsurl = {https://ui.adsabs.harvard.edu/abs/1992ApJS...81..387F}
}

@ARTICLE{2020A&A...642A..15K,
	author = {{Krucker}, S{\"a}m and {Hurford}, G.~J. and {Grimm}, O. and {K{\"o}gl}, S. and {Gr{\"o}belbauer}, H.-P. and {Etesi}, L. and {Casadei}, D. and {Csillaghy}, A. and {Benz}, A.~O. and {Arnold}, N.~G. and {Molendini}, F. and {Orleanski}, P. and {Schori}, D. and {Xiao}, H. and {Kuhar}, M. and {Hochmuth}, N. and {Felix}, S. and {Schramka}, F. and {Marcin}, S. and {Kobler}, S. and {Iseli}, L. and {Dreier}, M. and {Wiehl}, H.~J. and {Kleint}, L. and {Battaglia}, M. and {Lastufka}, E. and {Sathiapal}, H. and {Lapadula}, K. and {Bednarzik}, M. and {Birrer}, G. and {Stutz}, St. and {Wild}, Ch. and {Marone}, F. and {Skup}, K.~R. and {Cichocki}, A. and {Ber}, K. and {Rutkowski}, K. and {Bujwan}, W. and {Juchnikowski}, G. and {Winkler}, M. and {Darmetko}, M. and {Michalska}, M. and {Seweryn}, K. and {Bia{\l}ek}, A. and {Osica}, P. and {Sylwester}, J. and {Kowalinski}, M. and {{\'S}cis{\l}owski}, D. and {Siarkowski}, M. and {St{\k{e}}{\'s}licki}, M. and {Mrozek}, T. and {Podg{\'o}rski}, P. and {Meuris}, A. and {Limousin}, O. and {Gevin}, O. and {Le Mer}, I. and {Brun}, S. and {Strugarek}, A. and {Vilmer}, N. and {Musset}, S. and {Maksimovi{\'c}}, M. and {F{\'a}rn{\'\i}k}, F. and {Koz{\'a}{\v{c}}ek}, Z. and {Ka{\v{s}}parov{\'a}}, J. and {Mann}, G. and {{\"O}nel}, H. and {Warmuth}, A. and {Rendtel}, J. and {Anderson}, J. and {Bauer}, S. and {Dionies}, F. and {Paschke}, J. and {Pl{\"u}schke}, D. and {Woche}, M. and {Schuller}, F. and {Veronig}, A.~M. and {Dickson}, E.~C.~M. and {Gallagher}, P.~T. and {Maloney}, S.~A. and {Bloomfield}, D.~S. and {Piana}, M. and {Massone}, A.~M. and {Benvenuto}, F. and {Massa}, P. and {Schwartz}, R.~A. and {Dennis}, B.~R. and {van Beek}, H.~F. and {Rodr{\'\i}guez-Pacheco}, J. and {Lin}, R.~P.},
	title = "{The Spectrometer/Telescope for Imaging X-rays (STIX)}",
	journal = {\aap},
	year = 2020,
	month = oct,
	volume = {642},
	eid = {A15},
	pages = {A15},
	doi = {10.1051/0004-6361/201937362},
	adsurl = {https://ui.adsabs.harvard.edu/abs/2020A&A...642A..15K}
}

@ARTICLE{1963ApJ...137..945P,
	author = {{Pottasch}, Stuart R.},
	title = "{The Lower Solar Corona: Interpretation of the Ultraviolet Spectrum.}",
	journal = {\apj},
	year = 1963,
	month = apr,
	volume = {137},
	pages = {945},
	doi = {10.1086/147569},
	adsurl = {https://ui.adsabs.harvard.edu/abs/1963ApJ...137..945P}
}

@ARTICLE{1992PhyS...46..202F,
	author = {{Feldman}, Uri},
	title = "{REVIEW:  Elemental abundances in the upper solar atmosphere}",
	journal = {\physscr},
	year = 1992,
	month = sep,
	volume = {46},
	number = {3},
	pages = {202-220},
	doi = {10.1088/0031-8949/46/3/002},
	adsurl = {https://ui.adsabs.harvard.edu/abs/1992PhyS...46..202F}
}

@INPROCEEDINGS{1992ESASP.348..347S,
	author = {{Saba}, Julia L.~R. and {Strong}, Keith T.},
	title = "{Coronal abundances in solar active regions measured by the Solar Maximum Mission flat crystal spectrometer.}",
	booktitle = {Coronal Streamers, Coronal Loops, and Coronal and Solar Wind Composition},
	year = 1992,
	editor = {{Mattok}, C.},
	series = {ESA Special Publication},
	volume = {348},
	month = nov,
	pages = {347-350},
	adsurl = {https://ui.adsabs.harvard.edu/abs/1992ESASP.348..347S}
}

@ARTICLE{1995ApJ...440..884S,
	author = {{Sheeley}, Jr., N.~R.},
	title = "{A Volcanic Origin for High-FIP Material in the Solar Atmosphere}",
	journal = {\apj},
	year = 1995,
	month = feb,
	volume = {440},
	pages = {884},
	doi = {10.1086/175326},
	adsurl = {https://ui.adsabs.harvard.edu/abs/1995ApJ...440..884S}
}

@ARTICLE{2001ApJ...555..426W,
	author = {{Widing}, K.~G. and {Feldman}, U.},
	title = "{On the Rate of Abundance Modifications versus Time in Active Region Plasmas}",
	journal = {\apj},
	year = 2001,
	month = jul,
	volume = {555},
	number = {1},
	pages = {426-434},
	doi = {10.1086/321482},
	adsurl = {https://ui.adsabs.harvard.edu/abs/2001ApJ...555..426W}
}

@ARTICLE{2023ApJ...957...14S,
	author = {{Suarez}, Crisel and {Moore}, Christopher S.},
	title = "{Estimations of Elemental Abundances during Solar Flares Observed in Soft X-Rays by the MinXSS-1 CubeSat Mission}",
	journal = {\apj},
	year = 2023,
	month = nov,
	volume = {957},
	number = {1},
	eid = {14},
	pages = {14},
	doi = {10.3847/1538-4357/acf0c2},
	archivePrefix = {arXiv},
	eprint = {2308.16235},
	primaryClass = {astro-ph.SR},
	adsurl = {https://ui.adsabs.harvard.edu/abs/2023ApJ...957...14S}
}

@ARTICLE{2004ApJ...614.1063L,
	author = {{Laming}, J. Martin},
	title = "{A Unified Picture of the First Ionization Potential and Inverse First Ionization Potential Effects}",
	journal = {\apj},
	year = 2004,
	month = oct,
	volume = {614},
	number = {2},
	pages = {1063-1072},
	doi = {10.1086/423780},
	archivePrefix = {arXiv},
	eprint = {astro-ph/0405230},
	primaryClass = {astro-ph},
	adsurl = {https://ui.adsabs.harvard.edu/abs/2004ApJ...614.1063L}
}

@misc{zenodo_8409685,
	doi = {10.5281/zenodo.8409685},
	url = {https://doi.org/10.5281/zenodo.8409685},
	author = {{{Crameri}, C.}},
	title = {{Colormaps}},
	publisher = {Zenodo},
	year = {2023},
	month = {oct},
	note = {Accessed: 2023-10-21}
}

@article{ACTON199665,
	title = {Ancillary data services of NASA's Navigation and Ancillary Information Facility},
	journal = {Planetary and Space Science},
	volume = {44},
	number = {1},
	pages = {65-70},
	year = {1996},
	note = {Planetary data system},
	issn = {0032-0633},
	doi = {https://doi.org/10.1016/0032-0633(95)00107-7},
	url = {https://www.sciencedirect.com/science/article/pii/0032063395001077},
	author = {Charles H. Acton}
}

@article{ACTON20189,
	title = {A look towards the future in the handling of space science mission geometry},
	journal = {Planetary and Space Science},
	volume = {150},
	pages = {9-12},
	year = {2018},
	note = {Enabling Open and Interoperable Access to Planetary Science and Heliophysics Databases and Tools},
	issn = {0032-0633},
	doi = {https://doi.org/10.1016/j.pss.2017.02.013},
	url = {https://www.sciencedirect.com/science/article/pii/S0032063316303129},
	author = {Charles Acton and Nathaniel Bachman and Boris Semenov and Edward Wright}
}

@ARTICLE{2007SoPh..243...19C,
	author = {{Culhane}, J.~L. and {Harra}, L.~K. and {James}, A.~M. and {Al-Janabi}, K. and {Bradley}, L.~J. and {Chaudry}, R.~A. and {Rees}, K. and {Tandy}, J.~A. and {Thomas}, P. and {Whillock}, M.~C.~R. and {Winter}, B. and {Doschek}, G.~A. and {Korendyke}, C.~M. and {Brown}, C.~M. and {Myers}, S. and {Mariska}, J. and {Seely}, J. and {Lang}, J. and {Kent}, B.~J. and {Shaughnessy}, B.~M. and {Young}, P.~R. and {Simnett}, G.~M. and {Castelli}, C.~M. and {Mahmoud}, S. and {Mapson-Menard}, H. and {Probyn}, B.~J. and {Thomas}, R.~J. and {Davila}, J. and {Dere}, K. and {Windt}, D. and {Shea}, J. and {Hagood}, R. and {Moye}, R. and {Hara}, H. and {Watanabe}, T. and {Matsuzaki}, K. and {Kosugi}, T. and {Hansteen}, V. and {Wikstol}, {\O}.},
	title = "{The EUV Imaging Spectrometer for Hinode}",
	journal = {\solphys},
	year = 2007,
	month = jun,
	volume = {243},
	number = {1},
	pages = {19-61},
	doi = {10.1007/s01007-007-0293-1},
	adsurl = {https://ui.adsabs.harvard.edu/abs/2007SoPh..243...19C}
}

@ARTICLE{2025arXiv251002102M,
	author = {{Mondal}, Biswajit and {Winebarger}, Amy R.},
	title = "{Flare-Driven Plasma Dynamics and Elemental Abundance Redistribution}",
	journal = {arXiv e-prints},
	year = 2025,
	month = oct,
	eid = {arXiv:2510.02102},
	pages = {arXiv:2510.02102},
	doi = {10.48550/arXiv.2510.02102},
	archivePrefix = {arXiv},
	eprint = {2510.02102},
	primaryClass = {astro-ph.SR},
	adsurl = {https://ui.adsabs.harvard.edu/abs/2025arXiv251002102M}
}

@ARTICLE{2021A&C....3400449M,
	author = {{Mithun}, N.~P.~S. and {Vadawale}, S.~V. and {Patel}, A.~R. and {Shanmugam}, M. and {Chakrabarty}, D. and {Konar}, P. and {Sarvaiya}, T.~N. and {Padia}, G.~D. and {Sarkar}, A. and {Kumar}, P. and {Jangid}, P. and {Sarda}, A. and {Shah}, M.~S. and {Bhardwaj}, A.},
	title = "{Data processing software for Chandrayaan-2 Solar X-ray Monitor}",
	journal = {Astronomy and Computing},
	year = 2021,
	month = jan,
	volume = {34},
	eid = {100449},
	pages = {100449},
	doi = {10.1016/j.ascom.2021.100449},
	archivePrefix = {arXiv},
	eprint = {2007.11371},
	primaryClass = {astro-ph.IM},
	adsurl = {https://ui.adsabs.harvard.edu/abs/2021A&C....3400449M}
}

@ARTICLE{2020SoPh..295..139M,
	author = {{Mithun}, N.~P.~S. and {Vadawale}, Santosh V. and {Sarkar}, Aveek and {Shanmugam}, M. and {Patel}, Arpit R. and {Mondal}, Biswajit and {Joshi}, Bhuwan and {Janardhan}, P. and {Adalja}, Hiteshkumar L. and {Goyal}, Shiv Kumar and {Ladiya}, Tinkal and {Tiwari}, Neeraj Kumar and {Singh}, Nishant and {Kumar}, Sushil and {Tiwari}, Manoj K. and {Modi}, M.~H. and {Bhardwaj}, Anil},
	title = "{Solar X-Ray Monitor on Board the Chandrayaan-2 Orbiter: In-Flight Performance and Science Prospects}",
	journal = {\solphys},
	year = 2020,
	month = oct,
	volume = {295},
	number = {10},
	eid = {139},
	pages = {139},
	doi = {10.1007/s11207-020-01712-1},
	archivePrefix = {arXiv},
	eprint = {2009.09759},
	primaryClass = {astro-ph.SR},
	adsurl = {https://ui.adsabs.harvard.edu/abs/2020SoPh..295..139M}
}

@article{VADAWALE20142021,
	title = {Solar X-ray Monitor (XSM) on-board Chandrayaan-2 orbiter},
	journal = {Advances in Space Research},
	volume = {54},
	number = {10},
	pages = {2021-2028},
	year = {2014},
	note = {Lunar Science and Exploration},
	issn = {0273-1177},
	doi = {https://doi.org/10.1016/j.asr.2013.06.002},
	url = {https://www.sciencedirect.com/science/article/pii/S0273117713003438},
	author = {S.V. Vadawale and M. Shanmugam and Y.B. Acharya and A.R. Patel and S.K. Goyal and B. Shah and A.K. Hait and A. Patinge and D. Subrahmanyam}
}

@ARTICLE{2023ApJS..265...11D,
	author = {{Del Zanna}, Giulio and {Samra}, Jenna and {Monaghan}, Austin and {Madsen}, Chad and {Bryans}, Paul and {DeLuca}, Edward and {Mason}, Helen and {Berkey}, Ben and {de Wijn}, Alfred and {Rivera}, Yeimy J.},
	title = "{Coronal Densities, Temperatures, and Abundances during the 2019 Total Solar Eclipse: The Role of Multiwavelength Observations in Coronal Plasma Characterization}",
	journal = {\apjs},
	year = 2023,
	month = mar,
	volume = {265},
	number = {1},
	eid = {11},
	pages = {11},
	doi = {10.3847/1538-4365/acad68},
	archivePrefix = {arXiv},
	eprint = {2212.11889},
	primaryClass = {astro-ph.SR},
	adsurl = {https://ui.adsabs.harvard.edu/abs/2023ApJS..265...11D}
}

@ARTICLE{ZambranaPrado_2019,
	author = {{Zambrana Prado}, Natalia and {Buchlin}, {\'E}ric},
	title = "{Measuring relative abundances in the solar corona with optimised linear combinations of spectral lines}",
	journal = {\aap},
	year = 2019,
	month = dec,
	volume = {632},
	eid = {A20},
	pages = {A20},
	doi = {10.1051/0004-6361/201834735},
	archivePrefix = {arXiv},
	eprint = {1910.02886},
	primaryClass = {astro-ph.SR},
	adsurl = {https://ui.adsabs.harvard.edu/abs/2019A&A...632A..20Z}
}

@ARTICLE{Dere97,
	author = {{Dere}, K.~P. and {Landi}, E. and {Mason}, H.~E. and {Monsignori Fossi}, B.~C. and {Young}, P.~R.},
	title = "{CHIANTI - an atomic database for emission lines}",
	journal = {\aaps},
	year = 1997,
	month = oct,
	volume = {125},
	pages = {149-173},
	doi = {10.1051/aas:1997368},
	adsurl = {https://ui.adsabs.harvard.edu/abs/1997A&AS..125..149D}
}

@ARTICLE{Varesano_2024,
	author = {{Varesano}, T. and {Hassler}, D.~M. and {Zambrana Prado}, N. and {Plowman}, J. and {Del Zanna}, G. and {Parenti}, S. and {Mason}, H.~E. and {Giunta}, A. and {Auch{\`e}re}, F. and {Carlsson}, M. and {Fludra}, A. and {Peter}, H. and {M{\"u}ller}, D. and {Williams}, D. and {Aznar Cuadrado}, R. and {Barczynski}, K. and {Buchlin}, E. and {Caldwell}, M. and {Fredvik}, T. and {Grundy}, T. and {Guest}, S. and {Harra}, L. and {Janvier}, M. and {Kucera}, T. and {Leeks}, S. and {Schmutz}, W. and {Schuehle}, U. and {Sidher}, S. and {Teriaca}, L. and {Thompson}, W. and {Yardley}, S.~L.},
	title = "{SPICE connection mosaics to link the Sun's surface and the heliosphere}",
	journal = {\aap},
	year = 2024,
	month = may,
	volume = {685},
	eid = {A146},
	pages = {A146},
	doi = {10.1051/0004-6361/202347637},
	archivePrefix = {arXiv},
	eprint = {2308.01409},
	primaryClass = {astro-ph.SR},
	adsurl = {https://ui.adsabs.harvard.edu/abs/2024A&A...685A.146V}
}

@ARTICLE{To_2024,
	author = {{To}, Andy S.~H. and {Brooks}, David H. and {Imada}, Shinsuke and {French}, Ryan J. and {van Driel-Gesztelyi}, Lidia and {Baker}, Deborah and {Long}, David M. and {Ashfield}, IV, William and {Hayes}, Laura A.},
	title = "{Spatially resolved plasma composition evolution in a solar flare {\textendash} The effect of reconnection outflow}",
	journal = {\aap},
	year = 2024,
	month = nov,
	volume = {691},
	eid = {A95},
	pages = {A95},
	doi = {10.1051/0004-6361/202449246},
	archivePrefix = {arXiv},
	eprint = {2409.18188},
	primaryClass = {astro-ph.SR},
	adsurl = {https://ui.adsabs.harvard.edu/abs/2024A&A...691A..95T}
}

@ARTICLE{1976A&A....49..239C,
	author = {{Craig}, I.~J.~D. and {Brown}, J.~C.},
	title = "{Fundamental limitations of X-ray spectra as diagnostics of plasma temperature structure.}",
	journal = {\aap},
	year = 1976,
	month = jun,
	volume = {49},
	number = {2},
	pages = {239-250},
	adsurl = {https://ui.adsabs.harvard.edu/abs/1976A&A....49..239C}
}

@ARTICLE{1999A&A...348..286F,
	author = {{Fludra}, A. and {Schmelz}, J.~T.},
	title = "{The absolute coronal abundances of sulfur, calcium, and iron from Yohkoh-BCS flare spectra}",
	journal = {\aap},
	year = 1999,
	month = aug,
	volume = {348},
	pages = {286-294},
	adsurl = {https://ui.adsabs.harvard.edu/abs/1999A&A...348..286F}
}

@ARTICLE{2008A&A...488.1031C,
	author = {{Caffau}, E. and {Ludwig}, H.-G. and {Steffen}, M. and {Ayres}, T.~R. and {Bonifacio}, P. and {Cayrel}, R. and {Freytag}, B. and {Plez}, B.},
	title = "{The photospheric solar oxygen project. I. Abundance analysis of atomic lines and influence of atmospheric models}",
	journal = {\aap},
	year = 2008,
	month = sep,
	volume = {488},
	number = {3},
	pages = {1031-1046},
	doi = {10.1051/0004-6361:200809885},
	archivePrefix = {arXiv},
	eprint = {0805.4398},
	primaryClass = {astro-ph},
	adsurl = {https://ui.adsabs.harvard.edu/abs/2008A&A...488.1031C}
}

@ARTICLE{Plowman_2025,
	author = {{Plowman}, J.~E. and {Hassler}, D.~M. and {Molnar}, M.~E. and {Shrivastav}, A.~K. and {Varesano}, T. and {Auch{\`e}re}, F. and {Fludra}, A. and {Kucera}, T.~A. and {Wang}, T.~J. and {Zhu}, Y.},
	title = "{A new method of deriving Doppler velocities for Solar Orbiter SPICE}",
	journal = {\aap},
	year = 2026,
	month = feb,
	volume = {706},
	eid = {A171},
	pages = {A171},
	doi = {10.1051/0004-6361/202555756},
	archivePrefix = {arXiv},
	eprint = {2508.09121},
	primaryClass = {astro-ph.SR},
	adsurl = {https://ui.adsabs.harvard.edu/abs/2026A&A...706A.171P}
}

@ARTICLE{HMI_2012,
	author = {{Scherrer}, P.~H. and {Schou}, J. and {Bush}, R.~I. and {Kosovichev}, A.~G. and {Bogart}, R.~S. and {Hoeksema}, J.~T. and {Liu}, Y. and {Duvall}, T.~L. and {Zhao}, J. and {Title}, A.~M. and {Schrijver}, C.~J. and {Tarbell}, T.~D. and {Tomczyk}, S.},
	title = "{The Helioseismic and Magnetic Imager (HMI) Investigation for the Solar Dynamics Observatory (SDO)}",
	journal = {\solphys},
	year = 2012,
	month = jan,
	volume = {275},
	number = {1-2},
	pages = {207-227},
	doi = {10.1007/s11207-011-9834-2},
	adsurl = {https://ui.adsabs.harvard.edu/abs/2012SoPh..275..207S}
}

@ARTICLE{CROBAR_2021,
	author = {{Plowman}, Joseph},
	title = "{Three-dimensional Reconstruction of Coronal Plasma Properties from a Single Perspective}",
	journal = {\apj},
	year = 2021,
	month = dec,
	volume = {922},
	number = {2},
	eid = {109},
	pages = {109},
	doi = {10.3847/1538-4357/ac2664},
	archivePrefix = {arXiv},
	eprint = {2103.02028},
	primaryClass = {astro-ph.SR},
	adsurl = {https://ui.adsabs.harvard.edu/abs/2021ApJ...922..109P}
}

@ARTICLE{CROBAR_2023,
	author = {{Plowman}, Joseph},
	title = "{Validation and Testing of the CROBAR 3D Coronal Reconstruction Method with a MURaM Simulation}",
	journal = {\apj},
	year = 2023,
	month = apr,
	volume = {947},
	number = {1},
	eid = {5},
	pages = {5},
	doi = {10.3847/1538-4357/acbc71},
	archivePrefix = {arXiv},
	eprint = {2209.01753},
	primaryClass = {astro-ph.SR},
	adsurl = {https://ui.adsabs.harvard.edu/abs/2023ApJ...947....5P}
}

@ARTICLE{2022ApJ...939..112M,
	author = {{Mithun}, N.~P.~S. and {Vadawale}, Santosh V. and {Zanna}, Giulio Del and {Rao}, Yamini K. and {Joshi}, Bhuwan and {Sarkar}, Aveek and {{Mondal}, B} and {Janardhan}, P. and {Bhardwaj}, Anil and {Mason}, Helen E.},
	title = "{Soft X-Ray Spectral Diagnostics of Multithermal Plasma in Solar Flares with Chandrayaan-2 XSM}",
	journal = {The Astrophysical Journal},
	year = 2022,
	month = nov,
	volume = {939},
	number = {2},
	eid = {112},
	pages = {112},
	doi = {10.3847/1538-4357/ac98b4},
	archivePrefix = {arXiv},
	eprint = {2210.03364},
	primaryClass = {astro-ph.SR},
	adsurl = {https://ui.adsabs.harvard.edu/abs/2022ApJ...939..112M}
}

@ARTICLE{2025arXiv250814866M,
	author = {{Mondal}, Biswajit and {Winebarger}, Amy R. and {Athiray}, P.~S.},
	title = "{Abundance Diagnostics from Slitless Imaging Spectrometer: A Proof-of-Concept for MaGIXS-2}",
	journal = {Accepted in the Astrophysical Journal},
	year = 2025,
	month = aug,
	eid = {arXiv:2508.14866},
	pages = {arXiv:2508.14866},
	doi = {10.48550/arXiv.2508.14866},
	archivePrefix = {arXiv},
	eprint = {2508.14866},
	adsurl = {https://ui.adsabs.harvard.edu/abs/2025arXiv250814866M}
}

@ARTICLE{2012SoPh..275...17L,
	author = {{Lemen}, James R. and {Title}, Alan M. and {Akin}, David J. and {Boerner}, Paul F. and {Chou}, Catherine and {Drake}, Jerry F. and {Duncan}, Dexter W. and {Edwards}, Christopher G. and {Friedlaender}, Frank M. and {Heyman}, Gary F. and {Hurlburt}, Neal E. and {Katz}, Noah L. and {Kushner}, Gary D. and {Levay}, Michael and {Lindgren}, Russell W. and {Mathur}, Dnyanesh P. and {McFeaters}, Edward L. and {Mitchell}, Sarah and {Rehse}, Roger A. and {Schrijver}, Carolus J. and {Springer}, Larry A. and {Stern}, Robert A. and {Tarbell}, Theodore D. and {Wuelser}, Jean-Pierre and {Wolfson}, C. Jacob and {Yanari}, Carl and {Bookbinder}, Jay A. and {Cheimets}, Peter N. and {Caldwell}, David and {Deluca}, Edward E. and {Gates}, Richard and {Golub}, Leon and {Park}, Sang and {Podgorski}, William A. and {Bush}, Rock I. and {Scherrer}, Philip H. and {Gummin}, Mark A. and {Smith}, Peter and {Auker}, Gary and {Jerram}, Paul and {Pool}, Peter and {Soufli}, Regina and {Windt}, David L. and {Beardsley}, Sarah and {Clapp}, Matthew and {Lang}, James and {Waltham}, Nicholas},
	title = "{The Atmospheric Imaging Assembly (AIA) on the Solar Dynamics Observatory (SDO)}",
	journal = {\solphys},
	year = 2012,
	month = jan,
	volume = {275},
	number = {1-2},
	pages = {17-40},
	doi = {10.1007/s11207-011-9776-8},
	adsurl = {https://ui.adsabs.harvard.edu/abs/2012SoPh..275...17L}
}

@ARTICLE{Mithun_2021ExA....51...33M,
	author = {{Mithun}, N.~P.~S. and {Vadawale}, Santosh V. and {Shanmugam}, M. and {Patel}, Arpit R. and {Tiwari}, Neeraj Kumar and {Adalja}, Hiteshkumar L. and {Goyal}, Shiv Kumar and {Ladiya}, Tinkal and {Singh}, Nishant and {Kumar}, Sushil and {Tiwari}, Manoj K. and {Modi}, M.~H. and {Mondal}, Biswajit and {Sarkar}, Aveek and {Joshi}, Bhuwan and {Janardhan}, P. and {Bhardwaj}, Anil},
	title = "{Ground calibration of Solar X-ray Monitor on board the Chandrayaan-2 orbiter}",
	journal = {Experimental Astronomy},
	year = 2021,
	month = feb,
	volume = {51},
	number = {1},
	pages = {33-60},
	doi = {10.1007/s10686-020-09686-5},
	archivePrefix = {arXiv},
	eprint = {2007.07326},
	primaryClass = {astro-ph.IM},
	adsurl = {https://ui.adsabs.harvard.edu/abs/2021ExA....51...33M}
}

@ARTICLE{Mithun_2020SoPh..295..139M,
	author = {{Mithun}, N.~P.~S. and {Vadawale}, Santosh V. and {Sarkar}, Aveek and {Shanmugam}, M. and {Patel}, Arpit R. and {Mondal}, Biswajit and {Joshi}, Bhuwan and {Janardhan}, P. and {Adalja}, Hiteshkumar L. and {Goyal}, Shiv Kumar and {Ladiya}, Tinkal and {Tiwari}, Neeraj Kumar and {Singh}, Nishant and {Kumar}, Sushil and {Tiwari}, Manoj K. and {Modi}, M.~H. and {Bhardwaj}, Anil},
	title = "{Solar X-Ray Monitor on Board the Chandrayaan-2 Orbiter: In-Flight Performance and Science Prospects}",
	journal = {\solphys},
	year = 2020,
	month = oct,
	volume = {295},
	number = {10},
	eid = {139},
	pages = {139},
	doi = {10.1007/s11207-020-01712-1},
	archivePrefix = {arXiv},
	eprint = {2009.09759},
	primaryClass = {astro-ph.SR},
	adsurl = {https://ui.adsabs.harvard.edu/abs/2020SoPh..295..139M}
}

@ARTICLE{winebarger2012,
	author = {{Winebarger}, Amy R. and {Warren}, Harry P. and {Schmelz}, Joan T. and {Cirtain}, Jonathan and {Mulu-Moore}, Fana and {Golub}, Leon and {Kobayashi}, Ken},
	title = "{Defining the ``Blind Spot'' of Hinode EIS and XRT Temperature Measurements}",
	journal = {\apjl},
	year = 2012,
	month = feb,
	volume = {746},
	number = {2},
	eid = {L17},
	pages = {L17},
	doi = {10.1088/2041-8205/746/2/L17},
	adsurl = {https://ui.adsabs.harvard.edu/abs/2012ApJ...746L..17W}
}

@ARTICLE{fletcher2011,
	author = {{Fletcher}, L. and {Dennis}, B.~R. and {Hudson}, H.~S. and {Krucker}, S. and {Phillips}, K. and {Veronig}, A. and {Battaglia}, M. and {Bone}, L. and {Caspi}, A. and {Chen}, Q. and {Gallagher}, P. and {Grigis}, P.~T. and {Ji}, H. and {Liu}, W. and {Milligan}, R.~O. and {Temmer}, M.},
	title = "{An Observational Overview of Solar Flares}",
	journal = {\ssr},
	year = 2011,
	month = sep,
	volume = {159},
	number = {1-4},
	pages = {19-106},
	doi = {10.1007/s11214-010-9701-8},
	archivePrefix = {arXiv},
	eprint = {1109.5932},
	primaryClass = {astro-ph.SR},
	adsurl = {https://ui.adsabs.harvard.edu/abs/2011SSRv..159...19F}
}

@ARTICLE{holman2011,
	author = {{Holman}, G.~D. and {Aschwanden}, M.~J. and {Aurass}, H. and {Battaglia}, M. and {Grigis}, P.~C. and {Kontar}, E.~P. and {Liu}, W. and {Saint-Hilaire}, P. and {Zharkova}, V.~V.},
	title = "{Implications of X-ray Observations for Electron Acceleration and Propagation in Solar Flares}",
	journal = {\ssr},
	year = 2011,
	month = sep,
	volume = {159},
	number = {1-4},
	pages = {107-166},
	doi = {10.1007/s11214-010-9680-9},
	archivePrefix = {arXiv},
	eprint = {1109.6496},
	primaryClass = {astro-ph.SR},
	adsurl = {https://ui.adsabs.harvard.edu/abs/2011SSRv..159..107H}
}

@ARTICLE{Seaton2025,
	author = {{Seaton}, Daniel B. and {Downs}, Cooper and {Del Zanna}, Giulio and {West}, Matthew J. and {Thiemann}, Edward M.~B. and {Caspi}, Amir and {DeLuca}, Edward E. and {Golub}, Leon and {Mason}, James Paul and {Patel}, Ritesh and {Reeves}, Katharine K. and {Rivera}, Yeimy J. and {Savage}, Sabrina L.},
	title = "{Evidence of Extreme-ultraviolet Resonant Excitation in the Middle Corona during a Solar Flare}",
	journal = {\apj},
	year = 2025,
	month = may,
	volume = {985},
	number = {1},
	eid = {89},
	pages = {89},
	doi = {10.3847/1538-4357/adcab5},
	archivePrefix = {arXiv},
	eprint = {2504.08996},
	primaryClass = {astro-ph.SR},
	adsurl = {https://ui.adsabs.harvard.edu/abs/2025ApJ...985...89S}
}

@ARTICLE{2026ApJ...997..293P,
	author = {{Plowman}, Joseph and {Seaton}, Daniel B. and {Caspi}, Amir and {Hughes}, J. Marcus and {West}, Matthew J.},
	title = "{High-fidelity 3D Reconstruction of Solar Coronal Physics with the Updated CROBAR Method}",
	journal = {\apj},
	year = 2026,
	month = feb,
	volume = {997},
	number = {2},
	eid = {293},
	pages = {293},
	doi = {10.3847/1538-4357/ae0e11},
	archivePrefix = {arXiv},
	eprint = {2309.08053},
	primaryClass = {astro-ph.SR},
	adsurl = {https://ui.adsabs.harvard.edu/abs/2026ApJ...997..293P}
}

@ARTICLE{Vadawale_2021ApJ...912L..12V,
	author = {{Vadawale}, Santosh V. and {Mondal}, Biswajit and {Mithun}, N.~P.~S. and {Sarkar}, Aveek and {Janardhan}, P. and {Joshi}, Bhuwan and {Bhardwaj}, Anil and {Shanmugam}, M. and {Patel}, Arpit R. and {Adalja}, Hitesh Kumar L. and {Goyal}, Shiv Kumar and {Ladiya}, Tinkal and {Tiwari}, Neeraj Kumar and {Singh}, Nishant and {Kumar}, Sushil},
	title = "{Observations of the Quiet Sun during the Deepest Solar Minimum of the Past Century with Chandrayaan-2 XSM: Elemental Abundances in the Quiescent Corona}",
	journal = {\apjl},
	year = 2021,
	month = may,
	volume = {912},
	number = {1},
	eid = {L12},
	pages = {L12},
	doi = {10.3847/2041-8213/abf35d},
	archivePrefix = {arXiv},
	eprint = {2103.16643},
	primaryClass = {astro-ph.SR},
	adsurl = {https://ui.adsabs.harvard.edu/abs/2021ApJ...912L..12V}
}

@ARTICLE{2015ApJ...803...67D,
	author = {{Dennis}, Brian R. and {Phillips}, Kenneth J.~H. and {Schwartz}, Richard A. and {Tolbert}, Anne K. and {Starr}, Richard D. and {Nittler}, Larry R.},
	title = "{Solar Flare Element Abundances from the Solar Assembly for X-Rays (SAX) on MESSENGER}",
	journal = {\apj},
	year = 2015,
	month = apr,
	volume = {803},
	number = {2},
	eid = {67},
	pages = {67},
	doi = {10.1088/0004-637X/803/2/67},
	archivePrefix = {arXiv},
	eprint = {1503.01602},
	primaryClass = {astro-ph.SR},
	adsurl = {https://ui.adsabs.harvard.edu/abs/2015ApJ...803...67D}
}

@ARTICLE{1984Natur.310..665S,
	author = {{Sylwester}, J. and {Lemen}, J.~R. and {Mewe}, R.},
	title = "{Variation in observed coronal calcium abundance of X-ray flare plasmas}",
	journal = {\nat},
	year = 1984,
	month = aug,
	volume = {310},
	number = {5979},
	pages = {665-666},
	doi = {10.1038/310665a0},
	adsurl = {https://ui.adsabs.harvard.edu/abs/1984Natur.310..665S}
}

@ARTICLE{1995ApJ...447..936F,
	author = {{Fludra}, A. and {Schmelz}, J.~T.},
	title = "{Absolute Abundances of Flaring Coronal Plasma Derived from SMM Spectral Observations}",
	journal = {\apj},
	year = 1995,
	month = jul,
	volume = {447},
	pages = {936},
	doi = {10.1086/175931},
	adsurl = {https://ui.adsabs.harvard.edu/abs/1995ApJ...447..936F}
}

@ARTICLE{1981MNRAS.197...41V,
	author = {{Veck}, N.~J. and {Parkinson}, J.~H.},
	title = "{Solar abundances from X-ray flare observations}",
	journal = {\mnras},
	year = 1981,
	month = oct,
	volume = {197},
	pages = {41-55},
	doi = {10.1093/mnras/197.1.41},
	adsurl = {https://ui.adsabs.harvard.edu/abs/1981MNRAS.197...41V}
}

@ARTICLE{Zanna_2022ApJ...934..159D,
	author = {{Del Zanna}, G. and {Mondal}, B. and {Rao}, Y.~K. and {Mithun}, N.~P.~S. and {Vadawale}, S.~V. and {Reeves}, K.~K. and {Mason}, H.~E. and {Sarkar}, A. and {Janardhan}, P. and {Bhardwaj}, A.},
	title = "{Multiwavelength Observations by XSM, Hinode, and SDO of an Active Region. Chemical Abundances and Temperatures}",
	journal = {\apj},
	year = 2022,
	month = aug,
	volume = {934},
	number = {2},
	eid = {159},
	pages = {159},
	doi = {10.3847/1538-4357/ac7a9a},
	archivePrefix = {arXiv},
	eprint = {2207.06879},
	primaryClass = {astro-ph.SR},
	adsurl = {https://ui.adsabs.harvard.edu/abs/2022ApJ...934..159D}
}
\bibliographystyle{aasjournalv7}



\end{document}